\documentclass [%
superscriptaddress,
reprint,
 amsmath,amssymb,
 showkeys
]{revtex4-1}

\usepackage{graphicx}
\usepackage{dcolumn}
\usepackage{bm}
\usepackage{color}
\usepackage{float}
\usepackage[utf8]{inputenc}
\usepackage[mathlines]{lineno}
\usepackage{multirow}
\usepackage{hyperref}
\usepackage{amsmath}
\usepackage{xr}
\usepackage{enumerate}
\usepackage{array}
\usepackage{soul,xcolor}
\soulregister\cite7
\soulregister\ref7
\setstcolor{blue}
\newcommand{\AgI}{$\alpha$-AgI}

\newcommand{\wnws}{cm$^{-1}$ }

\newcommand{\app}{$\sim$}

\newcommand{\hmn}[1]{
  \ensuremath{\begingroup\setupHMN #1\endgroup}%
}

\newcommand{\setupHMN}{%
  \doHMN{-}{\HMNoverline}%
  \doHMN{*}{\HMNminverse}%
  \doHMN{i}{\infty}
}

\newcommand{\doHMN}[2]{%
  \begingroup\lccode`~=`#1
  \lowercase{\endgroup\let~}#2%
  \mathcode`#1="8000
}

\newcommand{\HMNminverse}[1]{\frac{#1}{m}}
\newcommand{\HMNoverline}[1]{\mkern1mu\overline{\mkern-1mu#1\mkern-1mu}\mkern1mu}

\newcolumntype{C}[1]{>{\centering\arraybackslash}m{#1}}

\begin{document}

\title{Anharmonic Host Lattice Dynamics Enable Fast Ion Conduction in Superionic AgI}

\author{Thomas M. Brenner}
\affiliation{Department of Materials and Interfaces, Weizmann Institute of Science, Rehovoth 76100, Israel}
\author{Christian Gehrmann}
\affiliation{Department of Physics, Technical University of Munich, 85748 Garching, Germany}
\author{Roman Korobko}
\affiliation{Department of Materials and Interfaces, Weizmann Institute of Science, Rehovoth 76100, Israel}
\author{Tsachi Livneh}
\affiliation{Department of Physics, Nuclear Research Center Negev, P.O. Box 9001, Beer Sheva 84190, Israel}
\author{David A. Egger}
\email{david.egger@tum.de}
\affiliation{Department of Physics, Technical University of Munich, 85748 Garching, Germany}
\author{Omer Yaffe}
\email{omer.yaffe@weizmann.ac.il}
\affiliation{Department of Materials and Interfaces, Weizmann Institute of Science, Rehovoth 76100, Israel}

\date{\today}

\begin{abstract}

Basic understanding of the driving forces of ion conduction in solids is critical to the development of new solid-state ion conductors.
Physical understanding of ion conduction is limited due to strong deviations from harmonic vibrational dynamics in these systems that are difficult to characterize experimentally and theoretically. 
We overcome this challenge in superionic AgI by combining THz-frequency Raman polarization-orientation measurements and \textit{ab-initio} molecular dynamics computations.
Our findings demonstrate clear signatures of strong coupling between the mobile ions and host lattice that are of importance to the diffusion process.
{We first derive a dynamic structural model from the Raman measurements that captures the simultaneous crystal-like and fluid-like properties of this fast-ion conductor.}
Then we show and discuss the importance of anharmonic relaxational motion that arises from the iodine host lattice by demonstrating its strong impact on ion conduction in superionic AgI.

\end{abstract}
%

\maketitle

\begin{figure}
	\centering
	\includegraphics[width=7.5 cm]{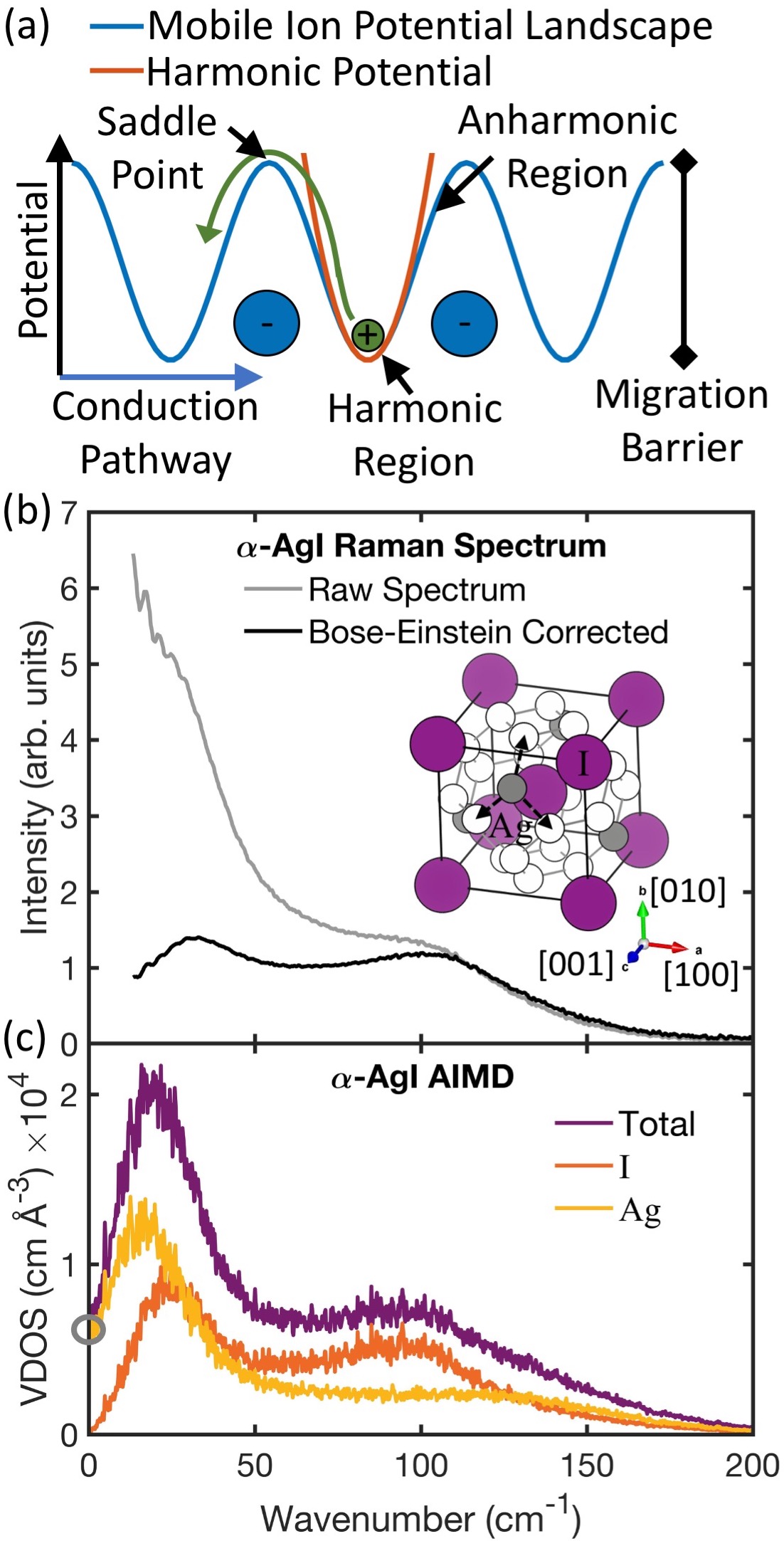}
	\caption{(a) Conventional ion transport mechanism: a mobile ion (green circle) hops (green arrow) across barriers in the potential landscape due to the host lattice (blue circles) and surrounding mobile ions. (b) \AgI\ Raman spectrum at 443 K. Inset: \AgI\ crystal structure, with I atoms (purple) located at BCC positions and Ag atoms (dark grey) hopping (black dashed arrows) between empty tetrahedral sites (white). (c) Vibrational density of states (VDOS) of \AgI\ calculated from AIMD at 500 K. The VDOS is decomposed into contributions from each atomic species. The gray circle on the y-axis marks the zero-frequency finite VDOS from which a diffusion coefficient of {$3.1\times10^{-5}$ cm$^{2}$/s} was calculated.}
	\label{RamanVDOS}
\end{figure}

Solid-state ion conductors (SSICs) are promising materials for replacing liquid ion conductors used in many important technological applications such as batteries, fuel cells, and supercapacitors, because they are typically non-flammable and could enable higher energy densities \cite{Manthiram:2017aa}. 
Long-range ionic diffusion in SSICs occurs \textit{via} thermally-activated site-to-site hopping of mobile ions in a host lattice of non-diffusing ions (Fig.~\ref{RamanVDOS}a).
Hopping occurs when mobile ions obtain sufficient thermal energy to overcome migration barriers in the potential energy surface. 
In traditional ionic transport models, the hopping frequency of the mobile ion is estimated using a harmonic phonon description
or included phenomenologically \cite{Rice:1958aa,Vineyard:1957aa,Almond:1987aa,Salamon:1979aa,Huggins:2009aa,Maier:2004aa}.
Therefore, these models cannot capture the effects that arise when the mobile ion enters the saddle point region between sites, an intrinsically anharmonic part of the potential energy surface (Fig.~\ref{RamanVDOS}a) \cite{Rice:1958aa,Vineyard:1957aa,Salamon:1979aa,Maier:2004aa}.
In this region, vibrations of the mobile ion and host lattice may become coupled, changing the potential energy surface of the mobile ion, and potentially modifying the migration barrier \cite{Rice:1958aa}. 
Accurately describing the atomic motions in SSICs therefore requires a departure from the harmonic phonon model of solids \cite{Rice:1958aa,Vineyard:1957aa,Niedziela:2019aa,Salamon:1979aa,Ding:2020aa}.
  
Recent computational and experimental work does indeed indicate that the host lattice impacts the ionic conductivity of SSICs in ways not yet captured by traditional theories \cite{Kraft:2017aa,Krauskopf:2017aa,Krauskopf:2018aa,Muy:2018aa,Kweon:2017aa,Adelstein:2016aa,Duchene:2019aa,Klerk:2016aa,Smith:2020aa}.
Correlations established between ionic conductivity and the physical characteristics of the host lattice, such as mechanical softness, indicate important unexplored coupling mechanisms \cite{Kraft:2017aa,Krauskopf:2017aa,Krauskopf:2018aa,Muy:2018aa}.  
Specific coupling effects, such as the `paddle-wheel' mechanism and chemical bond frustration, have been proposed \cite{Kweon:2017aa,Duchene:2019aa,Jansen:1991aa,Klerk:2016aa,Lunden:1988aa,Smith:2020aa,Adelstein:2016aa}.
However, major experimental and theoretical challenges hinder direct observation and identification of anharmonic effects arising from the host lattice.
Since this coupling cannot be captured in harmonic approximations of ionic conductivity, sophisticated analyses of molecular dynamics computations are required to detect it \cite{Kweon:2017aa,Adelstein:2016aa,Sagotra:2019aa}.
Experimental investigations of vibrational dynamics in SSICs must contend with the dynamic structural disorder that originates from hopping of mobile ions amongst partially populated sites (Fig.~\ref{RamanVDOS}b, inset).
In fast ion-conducting SSICs, this can result in the breakdown of the phonon quasiparticle picture \cite{Niedziela:2019aa,Ding:2020aa}, manifesting as broad features in their vibrational spectra.
These spectra are difficult to interpret because there is no analytical foundation describing the spectral properties of systems with dynamically broken translational symmetry \cite{Hanson:1975aa,Gallagher:1979aa,Alben:1977aa,Cazzanelli:1983aa,Niedziela:2019aa}.
Therefore, full interpretation of the vibrational spectrum of even \AgI, the best-understood and longest-studied superionic SSIC, still remains unresolved \cite{Hanson:1975aa,Gallagher:1979aa,Alben:1977aa,Cazzanelli:1983aa,Delaney:1977aa,Ushioda:1980aa}.
Overcoming these issues to provide a detailed account of vibrational coupling between mobile ion and host lattice is anticipated to provide useful insights into the role of significant anharmonicities in SSICs.     

In this letter, we demonstrate an approach to overcoming these challenges in \AgI.
We employ Raman polarization-orientation (PO) measurements to reveal unexpected signatures of order in the vibrational dynamics of \AgI. 
This finding enables the development of a novel microscopic description of the vibrational spectrum of this SSIC, providing a coherent interpretation of the origin of its features.
Our Raman measurements also reveal the presence of strongly anharmonic, relaxational atomic motions. 
Through \textit{ab-initio} molecular dynamics (AIMD) computations, we show that these relaxational motions involve the iodine host lattice and are coupled to Ag$^{+}$ diffusion. 
This discovery signifies that anharmonicity plays an important role in the functional properties of this material.

{$\alpha$-AgI exists at ambient pressure between 420-831~K.} \AgI\ has a cubic structure with time-average space group \hmn{Im-3m} where I atoms occupy BCC lattice sites and Ag atoms occupy interstitial tetrahedral sites (Fig.~\ref{RamanVDOS}b, inset) \cite{Cava:1977aa,Wright:1977aa,Hoshino:1977aa,Hoshino:1957aa,Cooper:1979aa,Boyce:1977aa,Wood:2006aa}.
There are 6 tetrahedral sites available per Ag atom, and the Ag$^{+}$ dynamically hop between these sites giving rise to the observed ion conductivity.
Since all Ag$^{+}$ participate in hopping and have a mean residence time of just 2-3 ps,\cite{Nemanich:1980aa,Funke:1976aa,Boyce:1977aa,Salamon:1979aa} hopping breaks the translational order of the Ag sublattice to below the unit cell level.

	{In materials with broken translational symmetry, optical vibrations are localized and lack well-defined wavevectors \cite{Shuker:1970aa}. 
	Consequently, the Raman spectrum of such a material is expected to have contributions from all modes, reflecting the vibrational density of states (VDOS) when corrected for the Bose-Einstein (BE) population factor \cite{Shuker:1970aa}.}
	The VDOS of \AgI\ can be calculated from AIMD (based on VASP \cite{Kresse:1996aa}, see Methods in supplementary material, SM \cite{SI}), since it is accurate in capturing the Ag$^+$ diffusion coefficient \cite{Kvist:1970aa} from the zero-frequency VDOS (gray circle, Fig.~\ref{RamanVDOS}c) \cite{Nitzan:2006aa}.
	Indeed, the BE-corrected Raman spectrum (Fig.~\ref{RamanVDOS}b) does reflect the AIMD-computed VDOS (Fig.~\ref{RamanVDOS}c), with both showing very broad features near 40~\wnws and 100~cm$^{-1}$. This observation is consistent with prior interpretations of the \AgI\ Raman spectrum, which have primarily ascribed the spectrum to disorder-induced scattering or a liquid-like scattering response \cite{Alben:1977aa,OSullivan:1991aa,Cazzanelli:1983aa,Delaney:1977aa,Ushioda:1980aa}.



\begin{figure*}
	\centering
	\includegraphics[width=\linewidth]{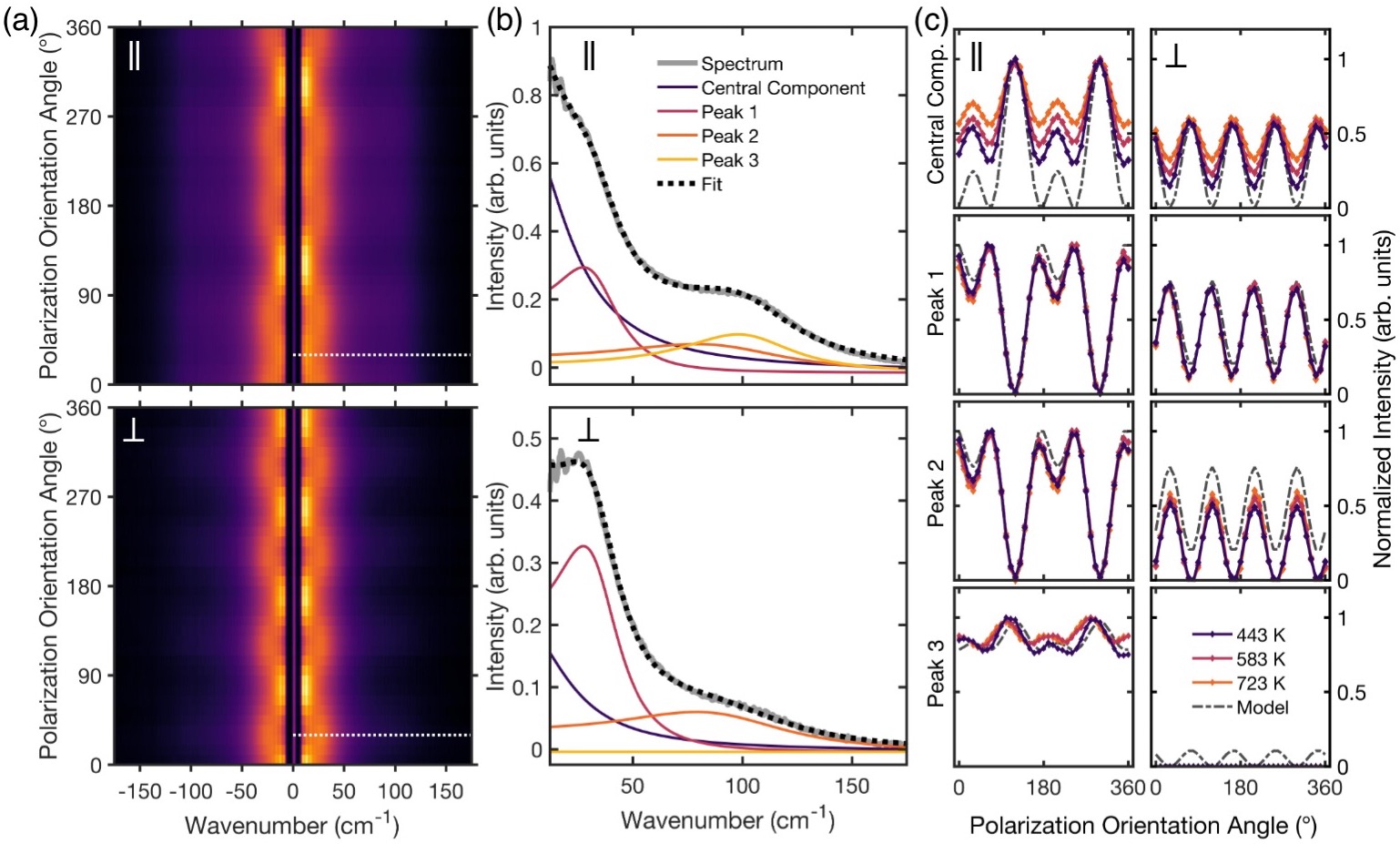}
	\caption{
		PO Raman measurements of \AgI. (a) PO intensity maps of Raman spectra taken at 443~K, at polarization orientation angles of 0-360$^\circ$ in parallel ($\parallel$) and perpendicular ($\bot$) polarization configurations, including both Stokes and anti-Stokes scattering. The region between -10 to 10 \wnws is cutoff by a notch filter. Spike-like features seen near 10 \wnws are due to interference effects from the notch filter. (b) Representative Raman spectra and their fit components and envelope, for both polarization configurations, taken from the dashed lines in part (a). (c) Plots of the normalized intensity coefficient versus polarization orientation angle for all four features, for each of the three temperatures measured. The expected response of the `local tetrahedral oscillator' model is also included (gray dashed line).
	}
	\label{Raman_results}
\end{figure*}

We now contrast the disorder-induced scattering picture with measurements from a method typically employed to characterize fully crystalline samples: PO Raman (Fig.~\ref{Raman_results}). In the PO method, the Raman spectrum of a single crystal is excited with polarized light whose orientation is rotated stepwise through 360$^\circ$ in the plane of the probed crystal face. The resulting spectral response is filtered for polarizations parallel and perpendicular to the excitation (see SM \cite{SI}) \cite{Mizoguchi:1989aa}.
This method probes the components of the Raman polarizability tensor, from which mode symmetries and consequently a structural model can be derived \cite{Hayes:2004aa,Mizoguchi:1989aa}.
Fig.~\ref{Raman_results}a shows intensity maps of the PO Raman response of an \AgI\ single crystal (preparation described in SM \cite{SI}) for the parallel and perpendicular configurations at 443~K.
{Additional measurements covering the stability range of $\alpha$-AgI (583~K and 723~K) are presented in Fig. S1 \cite{SI}.}
Remarkably, all of the Raman spectral features are found to exhibit clear periodicities in intensity as a function of orientation in both polarization configurations, which establishes specific symmetries for these vibrational features. This apparent crystal-like order in \AgI\ despite its broken translational symmetry is unique, since other materials with broken translational symmetry, such as liquids or amorphous solids, cannot show any modulation of intensity with orientation angle (Fig. S2).

The discovery of crystal-like periodicities in PO motivates development of a new structural model for \AgI. First, we identified and fitted each feature by employing the symmetry filtering effect of PO Raman (fit methodology in SM \cite{SI}); see Fig.~\ref{Raman_results}b for representative spectra and corresponding fits, and Table S1 for all fit parameters. 
{The fitting process revealed the characteristic PO intensity response for each Raman feature (Fig.~\ref{Raman_results}c), uncovering the four fundamental features of the} \AgI\ {vibrational spectrum:} one central component (fit by a Debye relaxor) and three broad peaks (fit by damped Lorentz oscillators \footnote{The extreme width of the peaks observed likely arises from a combination of lifetime and inhomogeneous broadening. The inhomogeneous broadening can arise from the many possible local environments of the `local tetrahedral oscillator', among other possible sources. Therefore, while we find the damped Lorentz oscillator is the only physically viable model at these low frequencies (see SM), the peak width likely does not reflect purely a vibrational lifetime.}) encompassing three distinct symmetries.

{Conventional Raman analysis fails to explain these observations:}
On the one hand, ascribing the Raman scattering to purely disorder-induced scattering as discussed above cannot explain the crystalline-like symmetries observed. 
On the other hand, standard crystal symmetry analysis \cite{Rousseau:1981aa} of the time-average \hmn{Im-3m} structure predicts only two Raman active features of two distinct symmetries (Table S2 and Table S3 \cite{SI}) and cannot capture the lower symmetry real-time structure which Raman is sensitive to.

\begin{figure}
	\centering
	\includegraphics[width=8 cm]{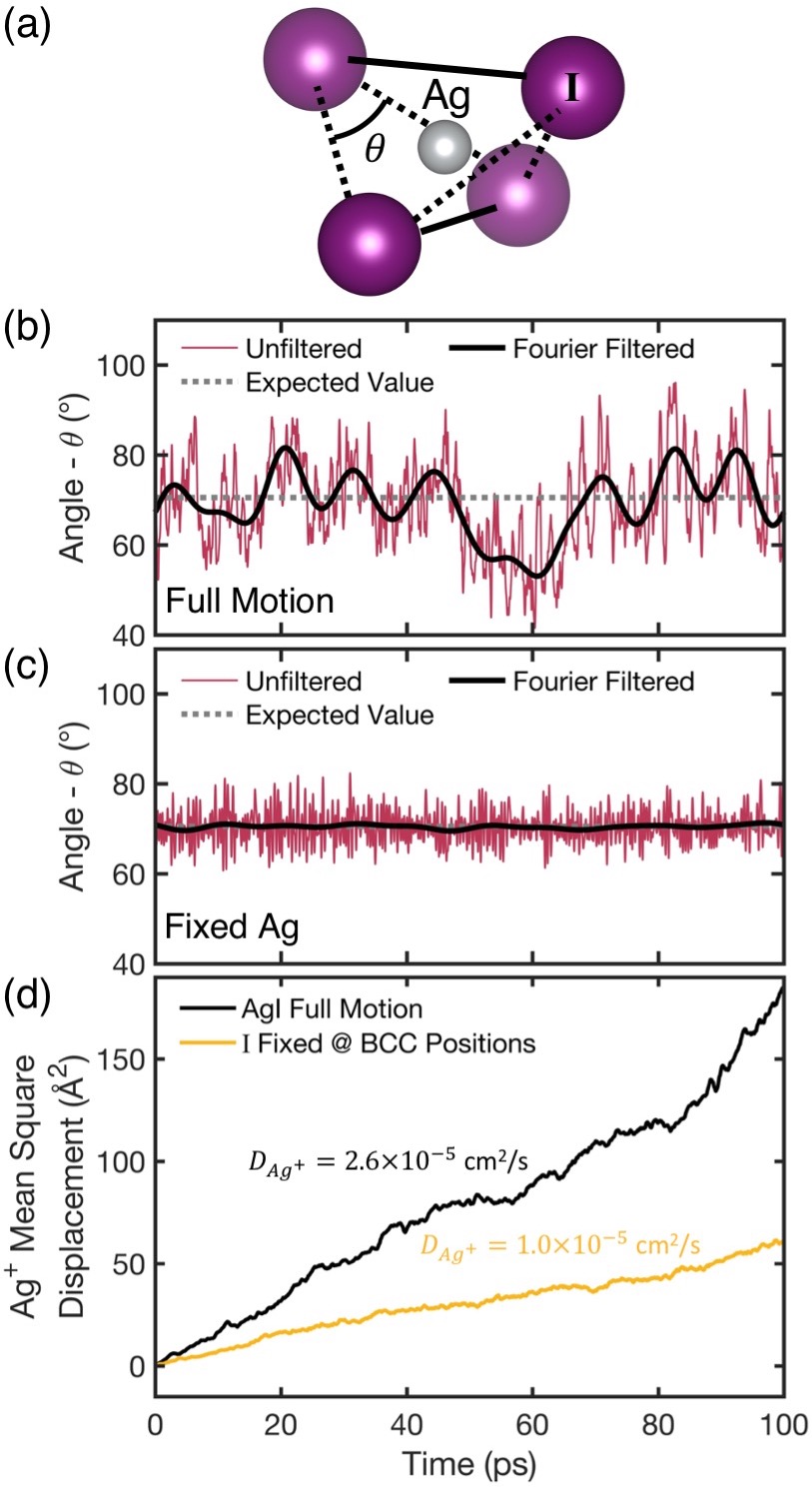}
	\caption{
		(a) Diagram of AgI$_4$ tetrahedral site of the BCC lattice, with the I-I-I bond angle $\theta$ analyzed in AIMD simulations indicated.
		Lines of the same texture have the same length.
		(b) Time series of $\theta$ for \AgI\ AIMD simulation at 500~K with all atomic motions allowed and (c) when Ag$^{+}$ is fixed to zero velocity at tetrahedral sites.
		(d) Mean square displacement as a function of time for Ag$^{+}$, and calculated diffusion coefficient, $D_{Ag^{+}}$, for the cases: all atomic motions allowed (black line),and I fixed at BCC lattice positions (yellow). 
}
	\label{MD_results}
\end{figure}

Instead, we propose a `local tetrahedral oscillator' model of vibrations that straddles the boundary between crystalline and disordered materials.
The critical innovation of our model is that the symmetry observed in the PO response is established locally rather than through the long-range order of a phonon model of vibrations.  
In our model, when an Ag$^+$ hops to a tetrahedral site of the BCC lattice, a transient, orientationally-ordered \footnote{`Orientational-order' denotes the integer number of tetrahedral ($D_{2d}$-symmetry) orientations allowed by the BCC lattice.}, local AgI$_4$ tetrahedral oscillator forms (Fig.~\ref{MD_results}a) whose symmetry is probed by our PO Raman measurements (details in SM, Table S3, and Figure S4 \cite{SI}).
This model predicts the number of Raman features we have observed in \AgI, their symmetry, and their PO response (Fig.~\ref{Raman_results}c, dashed line) and therefore provides a comprehensive picture of the vibrational dynamics in \AgI.

The most interesting feature identified from the PO Raman response is the central component (Fig.~\ref{Raman_results}), which is a categorical sign of anharmonic motion.
Central components are a manifestation of relaxational motion (i.e. motion that has no restoring force) in the frequency domain, characterized by a peak centered at zero frequency with a width that is determined by the characteristic relaxation time.
The central peak we have observed has a width of \app25~\wnws (Table S1 \cite{SI}) corresponding to a relaxation time of \app200 fs.
This is much faster than the polarizability changes directly due to ion hopping (2-3~ps relaxation time, see Fig. S4) \cite{Geisel:1977aa,Dieterich:1978aa,Nemanich:1980aa,Winterling:1977aa,Salamon:1979aa}, but similar to the time scale of relaxation in fluids (Fig. S5) \cite{Bucaro:1971aa,Fleury:1973aa,Bratos:1981aa,Dayan:1995aa,Mazzacurati:1990aa}. 
The physical origin and significance of this broader central component have not previously been considered \cite{Winterling:1977aa,Nemanich:1980aa}.

Our quantification of the PO response of this broad central component (Fig.~\ref{Raman_results}c) reveals that it is actually composed of two elements (Figure S6 \cite{SI}), one that is orientation-dependent (OD), and one that is orientation-independent (OI).
Because they have differing PO behavior, these elements must correspond to relaxational motions of differing physical origin.

The PO response of the OI element is distinctly ‘liquid-like’, a behavior not previously observed in a solid, to our knowledge. Like in a fluid, this element has no orientation dependence and is depolarized since it has a high ratio of perpendicular to parallel intensity (compare to central component of Figure S2 \cite{SI}). 
The OI element grows in prominence with increasing temperature in both parallel and perpendicular configurations (Fig. \ref{Raman_results}c, Fig. S6), but remains fully reversible and does not indicate, for instance, sample degradation effects.
The increasing dominance of this element with increasing temperature suggests a continuous change toward a more fluid-like condition throughout the stability region of the $\alpha$-phase.

The OD element of the central component is identified by its symmetry as one of the fundamental modes of the `local tetrahedral oscillator' model (Figure S6, Table S3 \cite{SI}).
If the system was fully harmonic, this element would present as a normal mode of finite frequency.
However, our PO Raman measurements have clearly established that the atomic motions involved are relaxational in nature.
To explain this, we hypothesize that the OD element of the central component arises from overdamping of a vibrational mode of the `local tetrahedral oscillator' through coupling to Ag$^+$ hopping.

Stimulated by these observations, we investigate via AIMD the relationship between iodine relaxational motion and Ag$^{+}$ diffusion. 
Our approach is to temporally and spatially resolve atomic motions in the AgI$_4$ tetrahedra of the `local tetrahedral oscillator' model.
Specifically, we investigate dynamically changing tetrahedral geometries by computing trajectories of the I-I-I bond angle $\theta$ (Fig.~\ref{MD_results}a) in one selected AgI$_4$ tetrahedron.
Any deviations of this bond angle from its expected value, i.e. the 70.5$^\circ$ angle of the BCC lattice, indicate distortions to the shape of the tetrahedron.
The red curve of Fig.~\ref{MD_results}b shows an AIMD trajectory of $\theta$, which displays very large angular fluctuations that appear on shorter but also longer time scales (e.g., see long-lived deviation at \app60 ps).
As discussed above, the most interesting aspect of the `local tetrahedral oscillator' model was the low-frequency relaxational motion giving rise to the central component in the PO Raman response.
To probe the atomic motions associated with it, we obtained trajectories of $\theta$ (black curve, Fig.~\ref{MD_results}b) retaining only the low frequency ($<5~\text{cm}^{-1}$) dynamics (see SM for further details \cite{SI}).
This frequency regime exclusively contains relaxational components that are not present in purely harmonic systems.
Fig.~\ref{MD_results}b demonstrates that in the low-frequency regime $\theta$ undergoes fluctuations of up to 15$^\circ$ from its expectation value, over a period of up to 10~ps, indicating large, long-lived rearrangements of the iodine tetrahedron that arise from anharmonic relaxation events.

A pair of AIMD-enabled \textit{gedankenexperiments} establishes the coupling between the iodine host lattice and Ag$^+$ motion. First, stipulating Ag$^{+}$ to be fixed with zero velocity at tetrahedral sites, we find that the large amplitude oscillations of $\theta$ at low frequency disappear (black curve, Fig.~\ref{MD_results}c).
This indicates that Ag$^{+}$ motion (either vibrational and/or hopping) is causally connected with relaxational motion of  the I lattice.
Second, to assess how I motion impacts Ag$^{+}$ diffusion, we consider the mean squared displacement of Ag$^{+}$ ions vs. time (Fig. 3d), the slope of which is proportional to the diffusion constant. Importantly, we find that freezing I motion impedes Ag$^{+}$ diffusion, which indicates that dynamic rearrangement of I is an important factor in ion diffusion, consistent with prior findings \cite{Schommers:1977aa,Hokazono:1984aa,Yokoyama:2003aa}. Further analysis shows that statically displaced I atoms, simulated at instantaneous configurations obtained as snapshots of the AIMD simulations, also lead to strongly reduced Ag$^{+}$ diffusion (Fig. S7).
These two \textit{gedankenexperiments} establish the strong interconnections between anharmonic relaxational I host lattice motions and the dynamics of Ag$^{+}$.

We suggest that a broad central component and corresponding relaxational motion of the host sublattice may be indicators of a good ion conductor. 
While a broader survey of materials is needed for confirmation, we do note that several other highly-conductive SSICs also show signs of a broad central component in their Raman response \cite{Nemanich:1980aa,Gallagher:1979aa,Chase:1976aa}.
Conversely, the much less conductive $\beta$-phase of AgI and non-ion conducting quasi-harmonic GaAs do not show a central component (Fig.~S5).
The proposed indicator would bring a physical understanding of microscopic dynamics to ionic conductor design that is not available from current empirical indicators \cite{Bachman:2016aa,Wang:2015aa}.
Moreover, this indicator explicitly includes information about the nature of the lattice dynamics, and specifically points out the need for anharmonic motion of the host lattice. 

In conclusion, using PO Raman measurements and AIMD calculations, we developed a new understanding of the vibrational dynamics of \AgI.
We introduced the `local tetrahedral oscillator' model in which the \AgI\ vibrational dynamics are localized to an AgI$_4$ tetrahedral configuration.
The success of our model demonstrates the value of a local picture of vibrational dynamics in materials with dynamic structural disorder.
Within that picture we have identified the presence of anharmonic relaxational motion in our Raman measurements and the AIMD simulations, which demonstrated an anharmonic relaxational component in the iodine host lattice.
Our AIMD computations indicate that iodine relaxation is causally connected to Ag$^{+}$ motion, and vice versa Ag$^{+}$ diffusion is strongly impacted by iodine vibrations.
These results show that anharmonic motion of the host lattice is a critical component of the ion transport process in \AgI.
We hope our findings revitalize the efforts to understand the deep connection between vibrational dynamics and ionic conductivity, and thereby may lead to new SSIC design principles.

\begin{acknowledgments}
We would like to thank Ishay Feldman (WIS) for performing X-Ray diffraction measurements, Lior Segev (WIS) for critical software development, and Claudio Cazorla (UNSW Sydney) and Brandon C. Wood (LLNL) for fruitful discussions. DAE acknowledges funding from: the Alexander von Humboldt Foundation within the framework of the Sofja Kovalevskaja Award, endowed by the German Federal Ministry of Education and Research; the Technical University of Munich - Institute for Advanced Study, funded by the German Excellence Initiative and the European Union Seventh Framework Programme under Grant Agreement No. 291763; the Deutsche Forschungsgemeinschaft (DFG, German Research Foundation) under Germany's Excellence Strategy - EXC 2089/1–390776260.
OY acknowledges funding from: the Benoziyo Endowment Fund, the Ilse Katz Institute, the Henry Chanoch Krenter Institute, the Soref New Scientists Start up Fund, Carolito Stiftung, the Abraham \& Sonia Rochlin Foundation, Erica A. Drake and Robert Drake, and the European Research Council.
\end{acknowledgments}
   
\bibliography{20191128_AgI_Paper_References}

\end{document}


\clearpage
\pagebreak
\widetext

\begin{center}
	\textbf{\Large Supplemental Materials: Anharmonic Host Lattice Dynamics Enable Fast Ion Conduction in Superionic AgI}
\end{center}

\setcounter{equation}{0}
\setcounter{figure}{0}
\setcounter{table}{0}
\setcounter{page}{1}
\makeatletter
\renewcommand{\theequation}{S\arabic{equation}}
\renewcommand{\thefigure}{S\arabic{figure}}
\renewcommand{\thetable}{S\arabic{table}}
\renewcommand{\bibnumfmt}[1]{[S#1]}
\renewcommand{\citenumfont}[1]{S#1}

\noindent{\textbf{\large Contents}}
\begin{enumerate}[I.]
	\item Supplementary Figures and Tables
	\item Methods
	\item The `Local Tetrahedral Oscillator' Model
\end{enumerate}


\section{Supplementary Figures and Tables}


\begin{figure*}[h]
	\centering
	\includegraphics[width=10 cm]{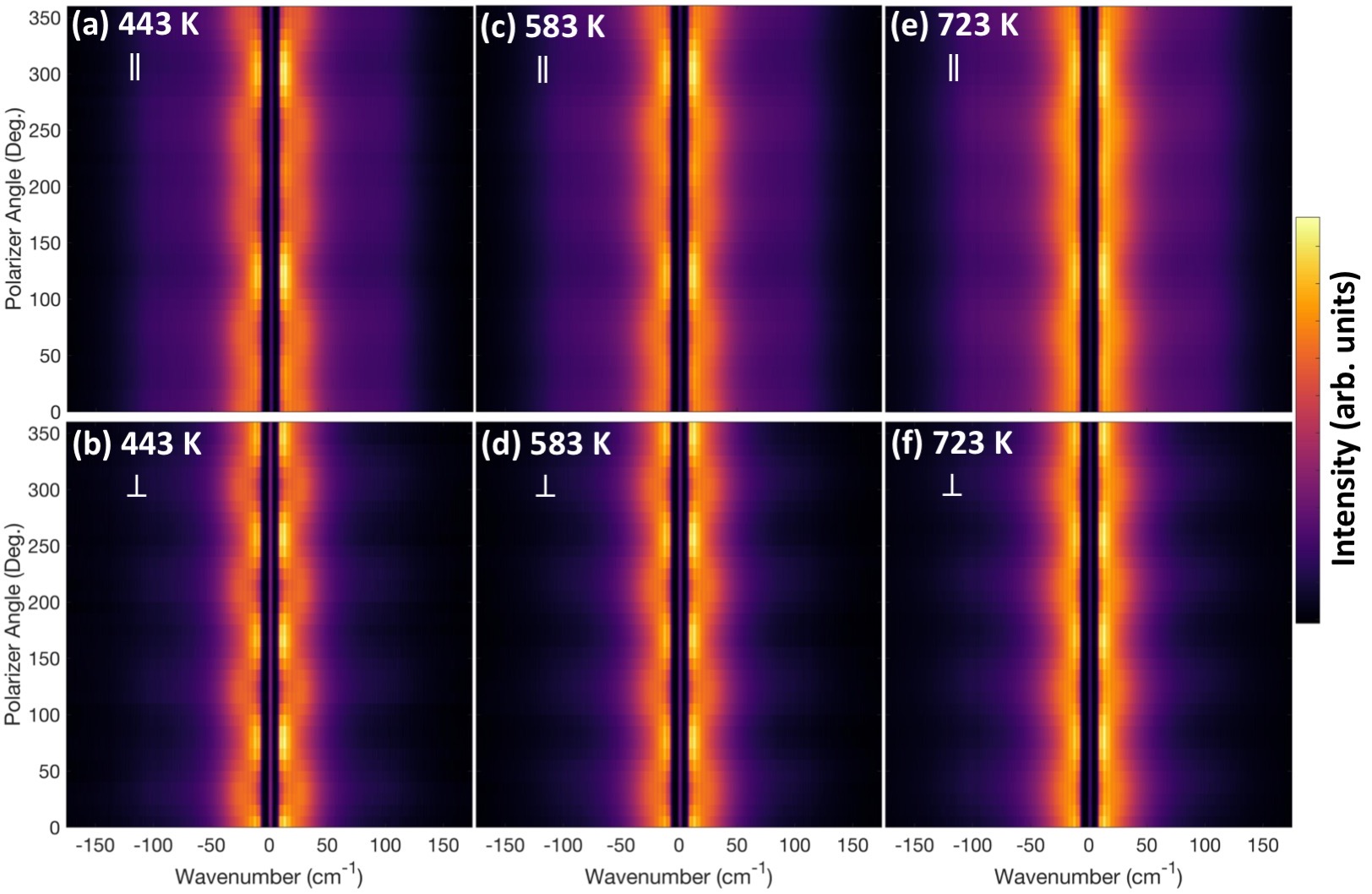}
	\caption{
		PO Raman measurements of an \AgI\ single crystal at temperatures of 443, 583, and 723 K. The top row gives measurements in the parallel ($ \parallel $) polarization configuration, bottom row in perpendicular ($ \perp $).
	}
	\label{PO_Temp}
\medskip
	\includegraphics[width=10 cm]{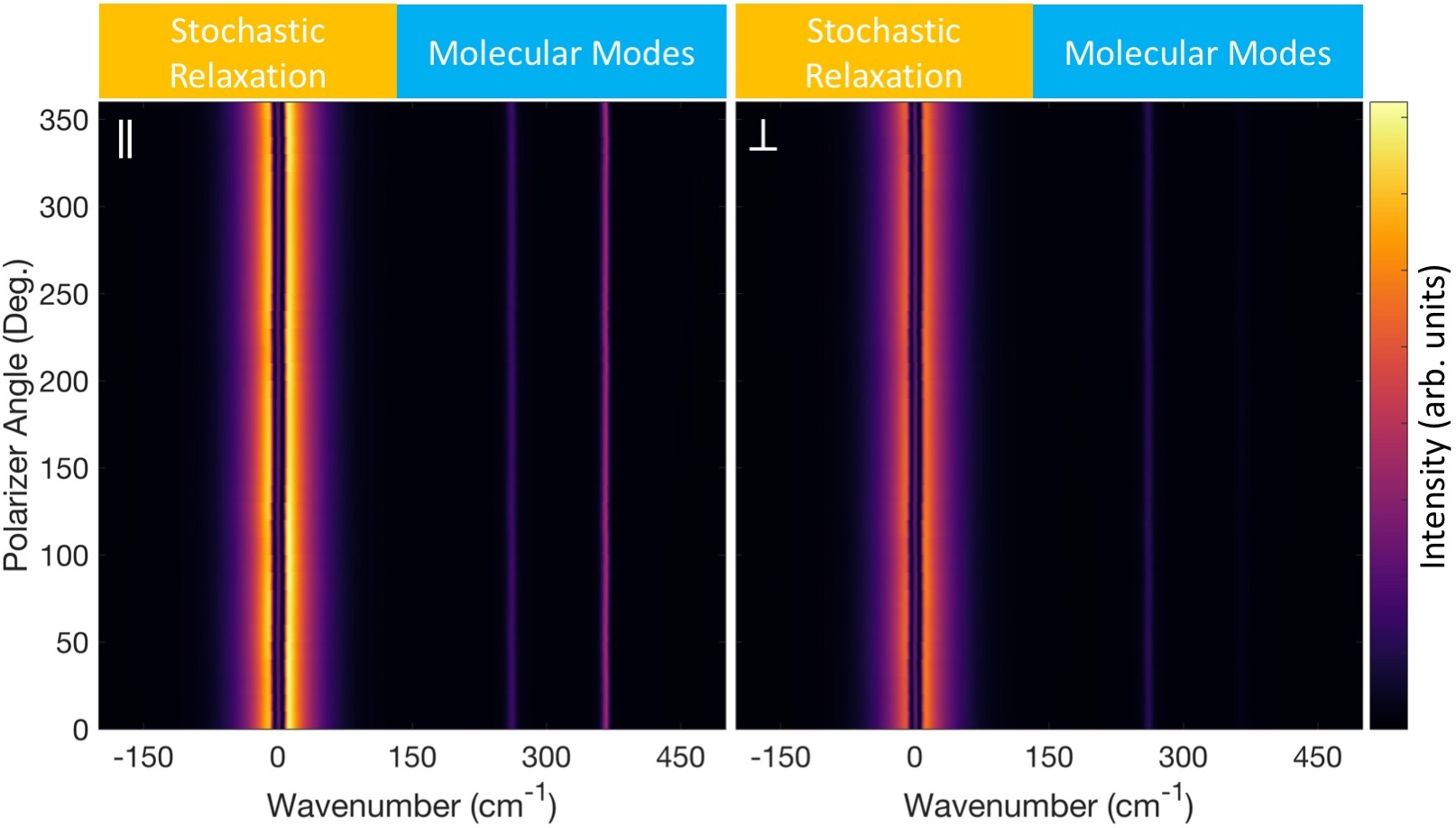}
	\caption{
		PO Raman measurement of chloroform (CHCl$_{3}$), a liquid at room temperature. There is no dependence of any feature on orientation angle due to the random orientation of the molecules. However, there are differences between the parallel and perpendicular analyzer configurations due to the symmetry of the molecular modes. There is an orientation-independent central component centered at 0 cm$^{-1}$ that indicates stochastic relaxation processes with no order. The central component has an intensity ratio of perpendicular to parallel close to 0.75, indicating it is depolarized, as is typical of a fluid \cite{Mariotto:1981aa}.
	}
	\label{CCl4}
\end{figure*}

\begin{table}[]
	\caption{Fit position and width parameters from fits to the PO Raman data sets for each temperature measured. These parameters were held constant for all orientation angles and both polarization configurations.}	
	\begin{ruledtabular}	
		\begin{tabular}{cccccccc}
			Fit Feature& 
			Central Component& 
			\multicolumn{2}{c}{Peak 1}& \multicolumn{2}{c}{Peak 2}& \multicolumn{2}{c}{Peak 3}\\
			\hline
			Temperature& Width& Position& Width&  Position& Width& Position& Width\\
			(K)& (cm$^{-1}$)& (cm$^{-1}$)& (cm$^{-1}$)& (cm$^{-1}$)& (cm$^{-1}$)& (cm$^{-1}$)& (cm$^{-1}$)\\
			\hline
			443& 23.0& 37.7& 37.0& 102.2& 95.7& 106.4& 61.4\\
			583& 24.3& 37.7& 37.0& 101.3& 115.7& 108.4& 64.1\\
			723& 26.5& 37.7& 37.0& 101.8& 137.7& 109.8& 67.5\\	
		\end{tabular}
	\end{ruledtabular}
	\label{Fit_Params}
\end{table}

\begin{table}[]
	\caption{Irreducible representation of the optical vibrational modes for each of the models tested. Row 1 gives the model type and the point group used to generate the irreducible representation. For the space group model, both the Hermann-Mauguin and the Mulliken notation are given for completeness. The irreducible representation of each model is given in row 2. Row 3 shows only the Raman active modes. Row 4 gives the number of Raman active modes. The total number of features that need to be explained is 4. Only the lower symmetry `local tetrahedral oscillator' can account for all features, as shown in Table~\ref{Assignments} below. Irreducible representations were determined employing the works of references \cite{Rousseau:1981aa,Nakamoto:2009aa}.}	
	\begin{ruledtabular}	
		\begin{tabular}{C{0.25\linewidth}C{0.3\linewidth}C{0.3\linewidth}}
			Model& \hmn{Im-3m} ($ O_{h}^{9} $) Space Group& AgI$_{4}$ $D_{2d}$ `Local Tetrahedral Oscillator'\\
			\hline
			Irreducible Representation of Optical Modes& $ A_{2g}+E_{g}+T_{2u}+T_{2g}+2T_{1u}+T_{1g} $& $ 2A_{1}+B_{1}+2B_{2}+2E $\\
			\hline
			Raman Active Modes& $ E_{g}+T_{2g} $& $ 2A_{1}+B_{1}+2B_{2}+2E $\\
			\hline
			Number of Raman Active Modes& 2& 7	
		\end{tabular}
	\end{ruledtabular}
	\label{Irreps}
\end{table}

\begin{table}[]
	\caption{Symmetry assignment of each fit component for each model tested. For the space group model, both the Hermann-Mauguin and the Mulliken notation are given for completeness. The mode symmetries are labeled with Mulliken symbols giving the irreducible representations of the point group. The peak frequency of each feature is provided.\\ $*$The number of Raman active modes in the $O_{h}^{9}$ model is insufficient to account for all features.\\ $\dagger$The central component has two elements (see Figure~\ref{NormCP} below). The symmetry assignment here is only for the polarization-dependent element. The liquid-like, polarization independent element has no symmetry to it.  
	}	
	\begin{ruledtabular}	
		\begin{tabular}{C{0.18\linewidth}C{0.18\linewidth}C{0.18\linewidth}C{0.18\linewidth}}
			Fit Component& \hmn{Im-3m} ($ O_{h}^{9} $) Space Group$^{*}$& AgI$_{4}$ $D_{2d}$ `Local Tetrahedral Oscillator'& Frequency (cm$^{-1}$)\\
			\hline
			Central Component$^{\dagger}$& $E_{g}$& $A_{1}+B_{1}$& 0\\
			Peak 1&	$T_{2g}$& $B_{2}+E$& 37.7\\
			Peak 2& ---& $B_{2}+E$& 102.2\\
			Peak 3& ---& $ A_{1} $& 106.4\\
		\end{tabular}
	\end{ruledtabular}
	\label{Assignments}
\end{table} 

\begin{figure*}[]
	\centering
	\includegraphics[width=11 cm]{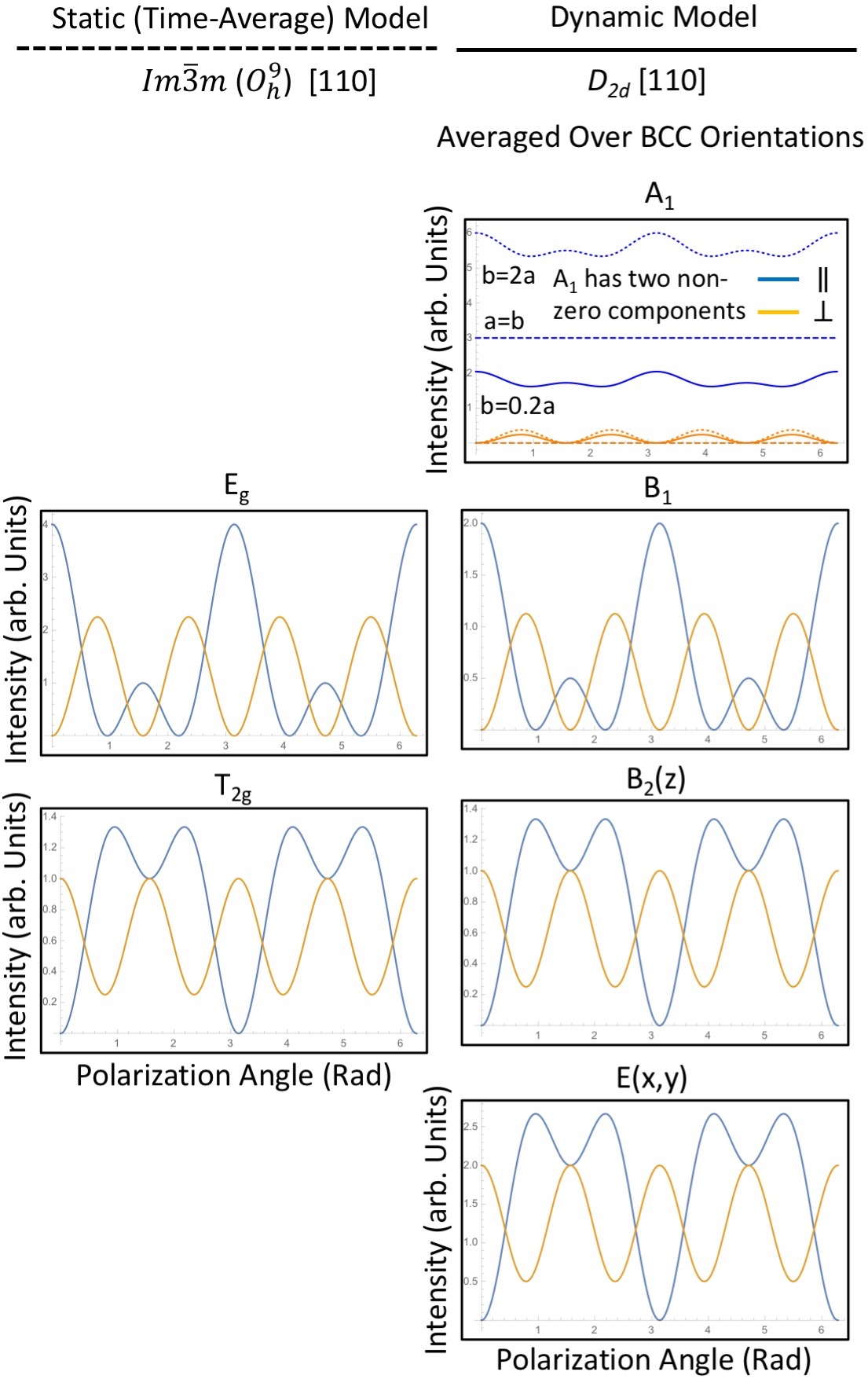}
	\caption{
		Expected orientation dependence of the Raman mode symmetries in each of the models tested, for light incident normal to the (110) plane of the BCC unit cell. 
		The details of these calculations are given in the methods section and the discussion of the `local tetrahedral oscillator' model below. 
		Both the parallel ($\parallel$, blue) and perpendicular ($\perp$, orange) orientation dependences are given. 
		Each plot is titled with the Mulliken symbol for the irreducible representation of the mode. 
		First column: polarization dependence expected for the modes of the static, time-average model, namely the space group \hmn{Im-3m} ($ O_{h}^{9} $). 
		Second column: polarization dependence expected for the modes of the `local tetrahedral oscillator' model of $ D_{2d} $ symmetry employed in the main text. 
		While modes with symmetry $ B_{1} $, $ B_{2} $, and $ E $ contain only one non-zero component in the Raman polarizability tensor, the $ A_{1} $ mode contains two non-zero components in its tensor. 
		Therefore, we have shown several possible ratios of the two tensor components ($ a $,$ b $): dashed line $ a=b $; solid line $ b=0.2a $; dotted line $ b=2a $. 
	}
	\label{POModels}
\end{figure*}

\begin{figure*}[]
	\centering
	\includegraphics[width=14 cm]{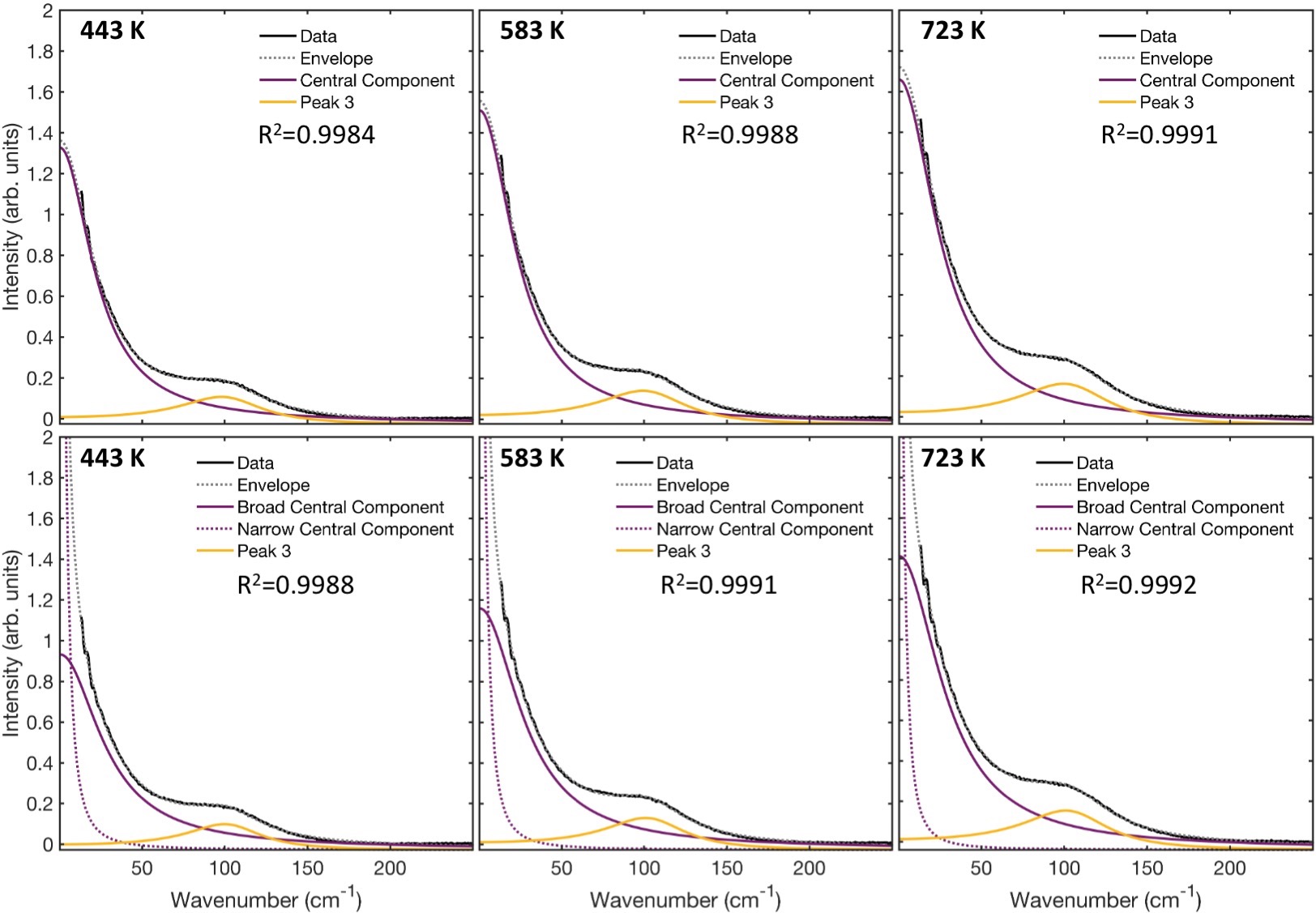}
	\caption{
		Comparison between fits with (bottom row) and without (top row) a narrow central component in addition to the broad central component. The narrow component was reported by Nemanich et al. as due to direct scattering from Ag$^{+}$ hopping events \cite{Nemanich:1980aa}. The fits demonstrate that the narrow component negligibly improves goodness-of-fit ($R^{2}$) and therefore contributes no additional information. While the narrow component likely contributes to the spectra, most of its intensity appears below the lower limit of our instrument (\app13 cm$^{-1}$), and this is likely why we cannot detect it in our measurements. The fitted spectra were taken at an orientation angle of 120$^\circ$ in the parallel configuration for all temperatures measured. At this polarization angle, only the central component and Peak 3 are present, giving the best information about the central component in the whole dataset. In the top row, the spectra are fit by a single damped Lorentz oscillator lineshape (Peak 3) and a single broad Debye relaxor lineshape. In the bottom row, an additional narrow (3 cm$^{-1}$) Debye relaxor lineshape is added. All parameters were left free to vary (position, width, and intensity), except for the width of the narrow central component. 
	}
	\label{TwoDebye}
\end{figure*}


\begin{figure*}[]
	\centering
	\includegraphics[width=9 cm]{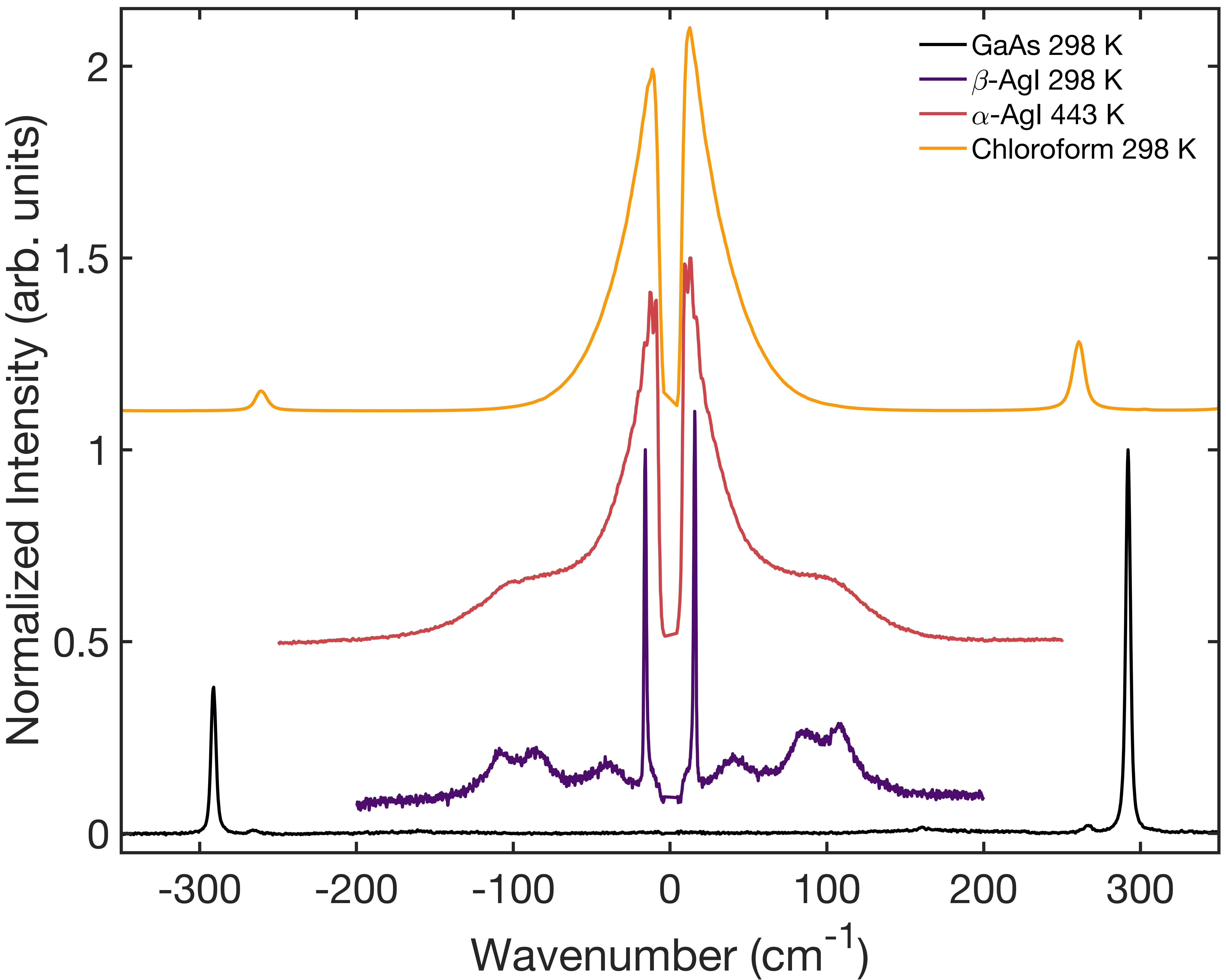}
	\caption{
		Raman spectra illustrating the low frequency response of several compounds. GaAs, which has no significant anharmonicity and has no ionic conductivity, shows no central component. $\beta$-AgI, which has very low ionic conductivity (\app$1\times10^{-5}$ S/cm) shows a very sharp optical mode at 17 cm$^{-1}$, but no central component. \AgI, with very high ionic conductivity (\app1 S/cm) shows a strong central peak, similar to the central peak displayed by liquids like chloroform (CHCl$_{3}$). The residual Rayleigh scattering in the region between -7 to 7 cm$^{-1}$ has been removed for clarity. 
	}
	\label{CPCompare}
\end{figure*}

\begin{figure*}[]
	\centering
	\includegraphics[width=9 cm]{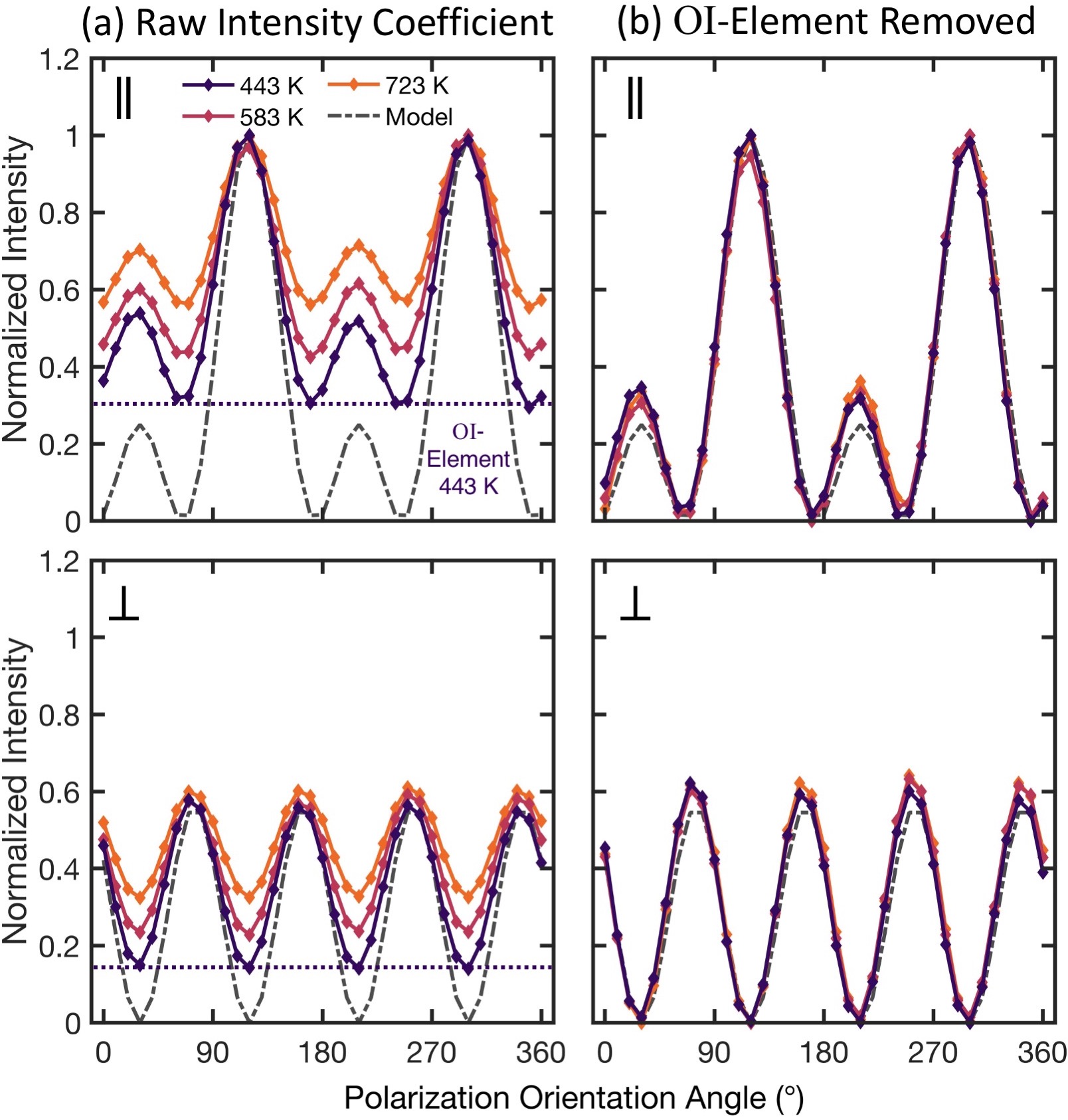}
	\caption{
		Decomposition of the broad (\app25 cm$^{-1}$) central component into orientation-independent (OI) and orientation-dependent (OD) elements. (a) shows the fitted raw intensity coefficients of the central component normalized to their maxima for the parallel ($\parallel$) and perpendicular ($\perp$) polarization configurations. The `local tetrahedral oscillator' model prediction for the OD-element is also shown. The intensity coefficients in both $\parallel$ and $\perp$ display a constant shift from the prediction of the model that increases as the temperature is increased. The constant shift is illustrated for 443~K by the dotted purple line. This constant intensity is identified as the OI-element of the central component. (b) shows the intensity coefficients after removing the OI-element in comparison to the predicted intensity response of the `local tetrahedral oscillator' model. From the strong agreement with the `local tetrahedral oscillator' model, we identify this remaining intensity as the OD-element and recognize that it is one of the fundamental modes of the model, as discussed in the main text.
	}
\label{NormCP}
\end{figure*}

\begin{figure*}[]
	\centering
	\includegraphics[width=10 cm]{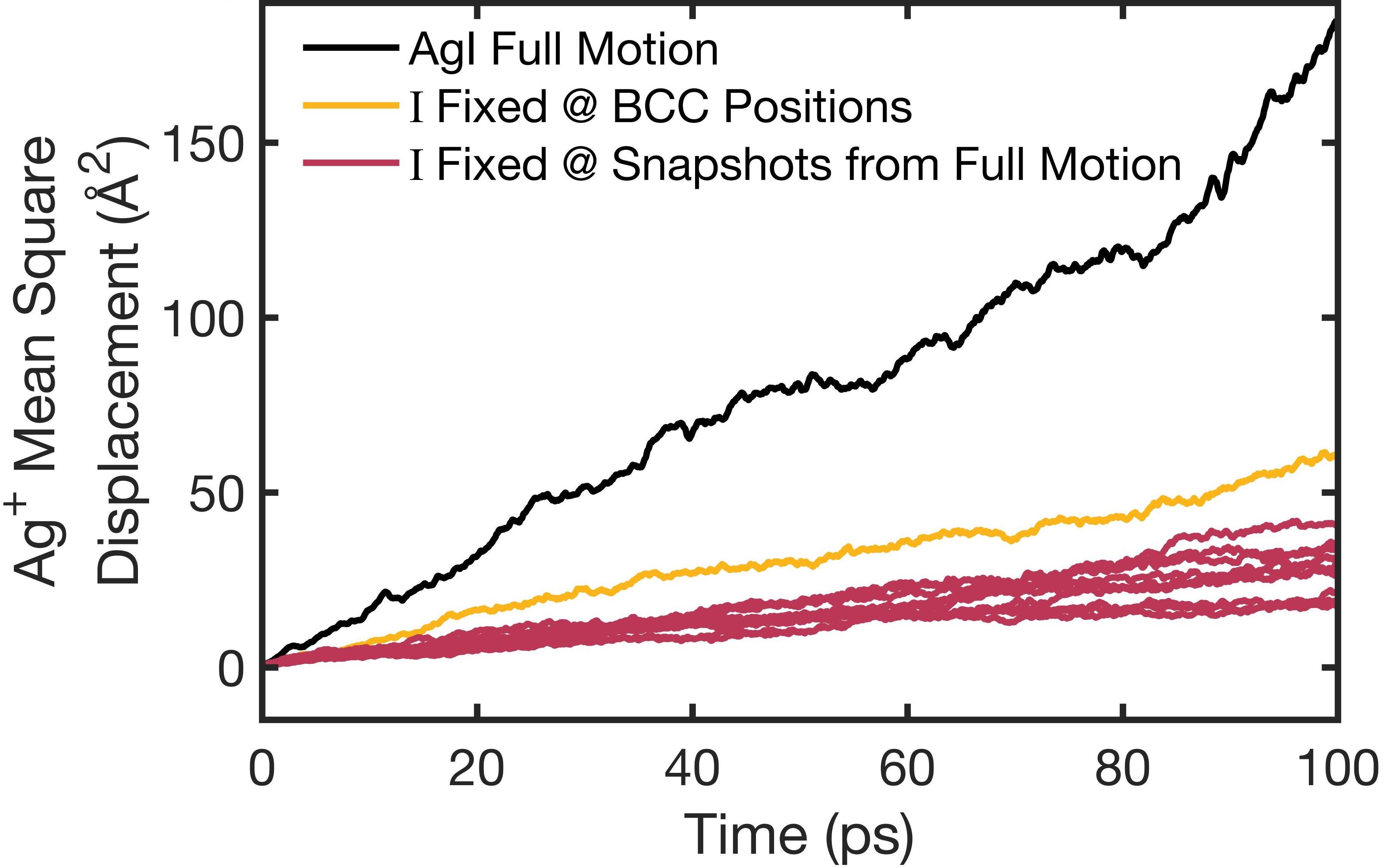}
	\caption{Mean squared displacement of Ag$^{+}$ ions vs. time, the slope of which is proportional to the diffusion constant. We find that freezing I motion (yellow and red lines) impedes Ag$^{+}$ diffusion compared to full motion (black lines), which indicates that dynamic rearrangement of I is an important factor in ion diffusion. Further insight into the role of dynamic I rearrangements for Ag$^{+}$ diffusion is established when comparing diffusion constants for two relevant scenarios: that of I atoms being fixed at BCC lattice positions (yellow line) versus if they are first allowed to thermally equilibrate with full motion, and then frozen at instantaneous configurations obtained as snapshots of the AIMD simulations (red lines). Interestingly, the latter scenario results in a smaller Ag$^{+}$ diffusion constant. This can be explained considering that the transition barriers for Ag$^{+}$ hopping in the BCC scenario are all equal, whereas in the instantaneous scenarios the barriers take on a distribution of heights in which the higher barriers are limiting Ag$^{+}$ diffusion.
	}
	\label{Diffusion}
\end{figure*}

\clearpage
\pagebreak

\section{Methods}
\noindent{\textit{Preparation of $ \beta $-AgI Single Crystals}}

Crystals were prepared according to a published procedure \cite{Helmholz:1935aa}. 
The following was performed under ambient atmosphere, in regular lab lighting conditions: 57 wt.\% HI (unstabilized, 99.99\%, Sigma Aldrich) was diluted with an equal volume of deionized H$ _{2} $O. 
2.5g of AgI powder (99.9\%, Alfa Aesar) were dissolved in 20 mL of the diluted HI by gentle heating (0.125 g/mL). 
Glass vials containing 5 mL of solution were prepared, and 2.5 mL of methanol was added to each. 
The vials were kept with caps loosely placed on top. 
The vials were kept in an undisturbed area, in the dark for approx. 1 month while crystals precipitated. 
Upon removal, the crystals were rinsed with methanol and then soaked in methanol for several days to remove HI inclusions. 
Only crystals showing the expected hexagonal pyramid habit were used for measurement. 
The pyramid basal plane was determined to be (002) by x-ray diffraction pole figures, as expected from literature \cite{Helmholz:1935aa}.
\medskip

\noindent{\textit{Preparation of \AgI\ Crystals from $\beta$-AgI Single Crystals}}

\AgI\ crystals were prepared from $\beta$-AgI single crystals according to the procedure of Mills et al. \cite{Mills:1970aa}. A hexagonal pyramid-shaped $\beta$-AgI single crystal was cleaved with a scalpel parallel to the base to form a flat clean surface. $\beta$-AgI readily cleaves along this plane (the (002) plane). The cleaved crystal was dropped onto a heated (523K) microscope stage (Linkam THMS600) with its base parallel to the stage. It is critical to drop the crystal onto its base, and not any other face, and to keep the crystal above 423K at all times. This procedure produces crystals oriented with the (110) plane parallel to the microscope stage, i.e. (002) $\rightarrow$ (110), as verified below (see section on self-consistent orientation determination) and expected from literature \cite{Mills:1970aa}.
\medskip

\noindent{\textit{Temperature-Dependent Polarization-Orientation (PO) Raman Measurement}}

\begin{figure*}
	\centering
	\includegraphics[width=9 cm]{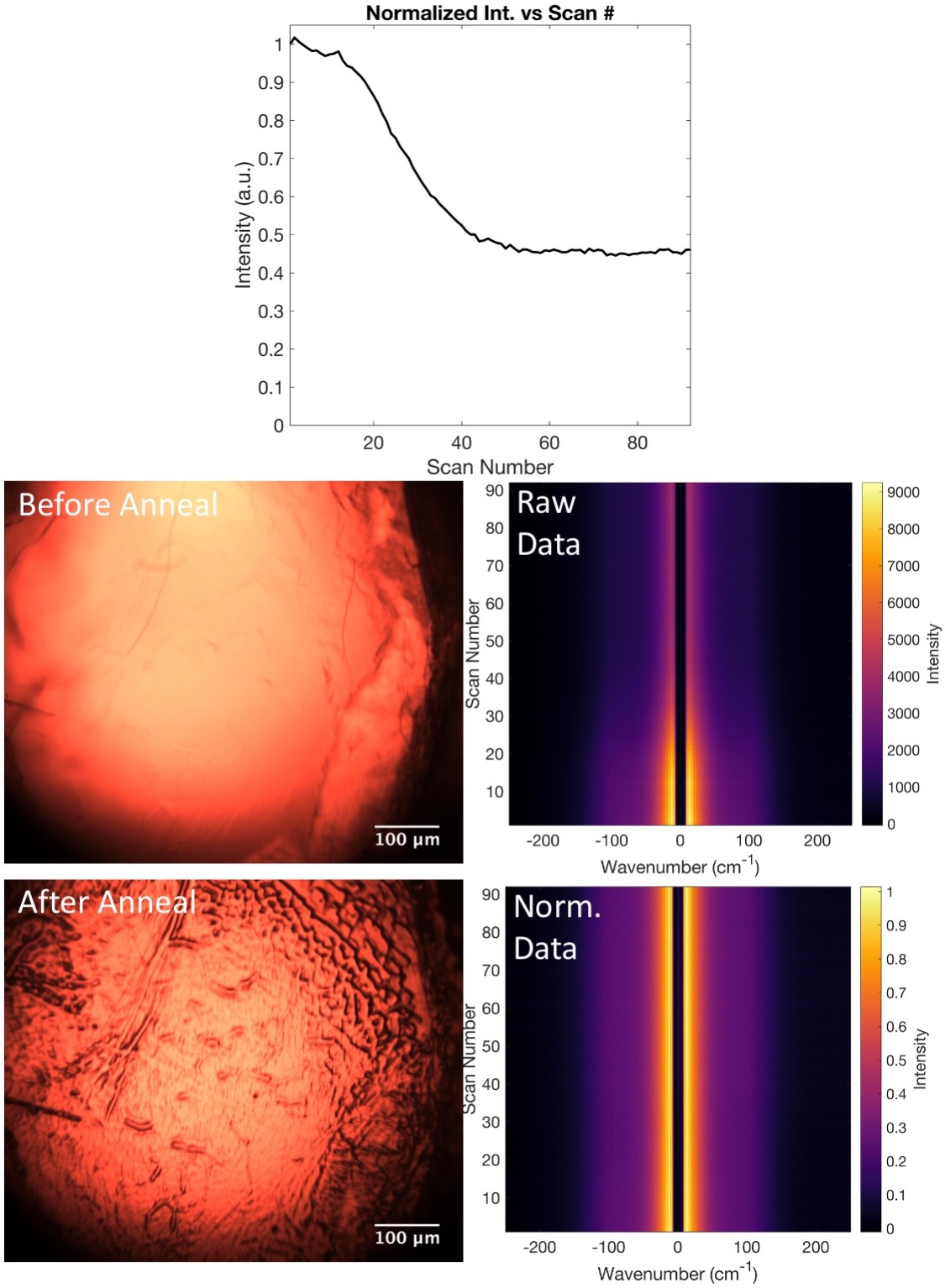}
	\caption{
		Time dependence of $\alpha$-AgI single crystal morphology and Raman intensity at 723~K. Raman measurements were taken continuously over ~15 hrs while holding the sample at 723~K. The Raman intensity steadily drops over time, and then stabilizes (top panel). 
		This process takes 6-8 hours. 
		The microscope images show the crystal morphology changes that occured during this time. 
		The Raman intensity maps show the change in Raman spectra with time, measured in a single polarization/orientation configuration. 
		The raw data (Raw Data) shows the drop and subsequent stabilization of intensity. 
		By normalizing to the maximum intensity value of each individual spectrum (Norm. Data) it becomes clear that the lineshape does not change during this intensity drop. We conclude that the drop in Raman intensity is due to the morphological changes taking place at the sample surface. After stabilization, the PO Raman measurement can be taken. 
	}
	\label{Morph}
\end{figure*}

PO employs polarized light to probe the components of the Raman polarizability tensor in a single crystal, providing identification of mode symmetries \cite{Mizoguchi:1989aa}. The measurements were performed on a Zeiss Axio Vario Scope.A1 microscope with a 20x NIR objective in backscatter configuration with the excitation source impinging perpendicular to the (110) plane of \AgI\ (as verified self-consistently from the PO measurements, and expected from literature, see below) \cite{Mills:1970aa}. The sample was enclosed in a Linkam THMS 600 microscope heating stage under N2 purge. The sample temperature was controlled via the Linkam. We note that care must be taken with PO Raman measurements of $\alpha$-AgI at high temperatures. We observed microscopic morphological changes in the sample over time at 723K that correlated to a drop in Raman intensity, without lineshape change (Figure~\ref{Morph}), followed by stabilization. PO Raman could only be performed after this process completed. This may explain the drop in Raman intensity with increasing temperature reported in previous literature \cite{Fontana:1980aa}.

\begin{figure*}
	\centering
	\includegraphics[width=10 cm]{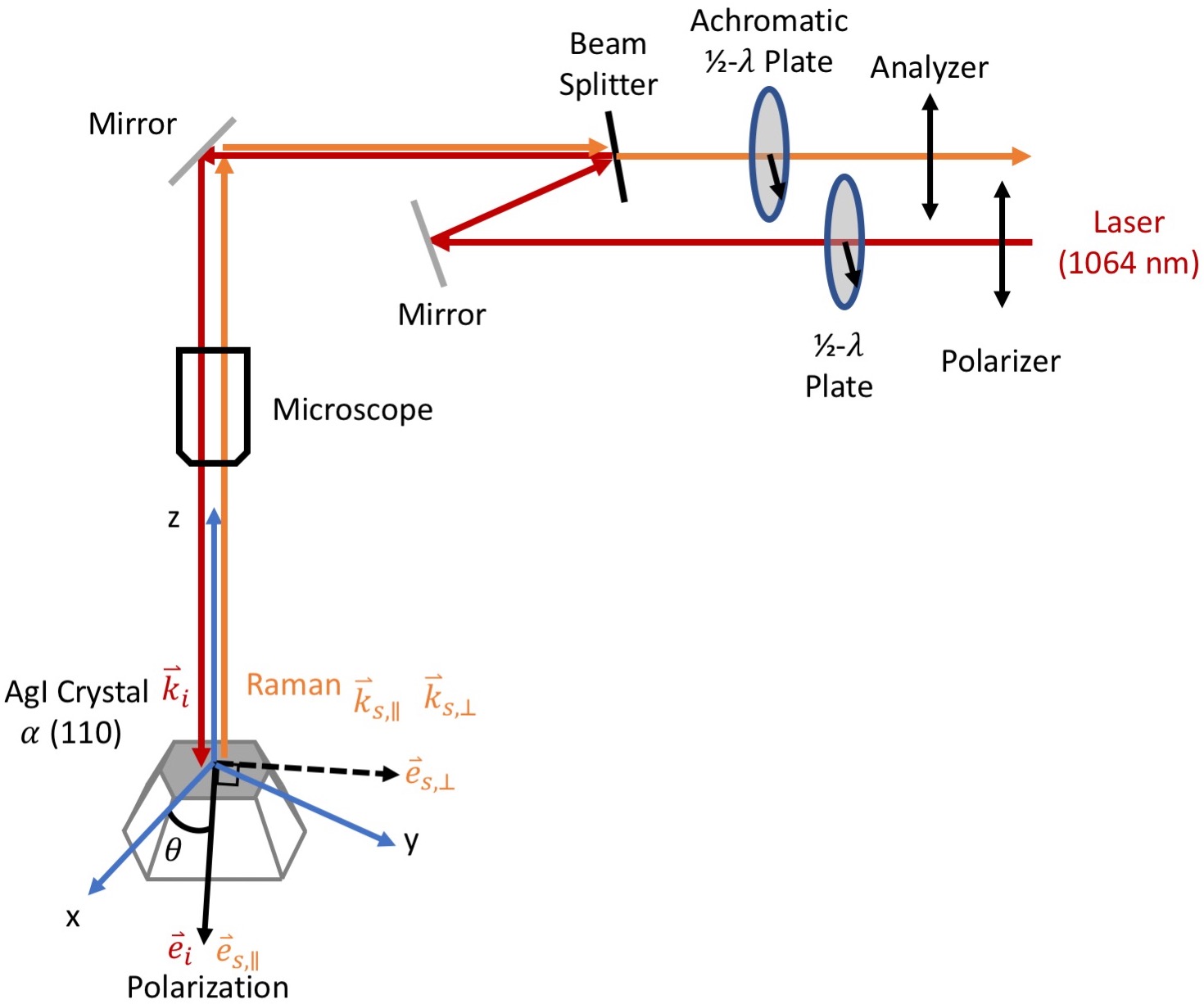}
	\caption{
		Schematic for PO Raman measurement in backscatter configuration. 1064 nm laser light is polarized and directed at normal incidence ($\vec{k}_{i}$) to the (110) plane of an $\alpha$-AgI crystal, with its electric field polarized with an angle $\theta$ relative to the x-axis ($\vec{e}_{i}$). The x-y plane is equivalent to the (110) plane. Backscattered light ($\vec{k}_{s,\parallel}$,$\vec{k}_{s,\perp}$) is polarization-filtered by another polarizer (analyzer) for polarizations parallel ($\vec{e}_{s,\parallel}$) and perpendicular ($\vec{e}_{s,\perp}$) to the incident polarization, $\vec{e}_{i}$. During a PO Raman experiment, the polarization angle $\theta$ of $\vec{e}_{i}$ is rotated stepwise through 360$^{\circ}$ and spectra are recorded in the parallel ($\parallel$) and perpendicular ($\perp$) analyzer configurations for each step. 
	}
	\label{ExpSetup}
\end{figure*}

The Raman crystallography measurement setup is shown in Figure~\ref{ExpSetup}. 

A 1064 nm laser (Coherent Mephisto) was used for the Raman excitation. This proved critical as shorter wavelength lasers caused a photographic effect during measurement, ruining the long-time scale measurements required for Raman crystallography. The laser was filtered for any amplified spontaneous emission (ASE) using volume holographic ASE filters (Ondax). The beam shape was optimized by a pinhole spatial filter. The laser was polarization filtered by a calcite polarizer (Thorlabs), and the polarization was rotated with respect to the sample with a monochromatic half-wave plate (Thorlabs). The polarized beam is directed into the microscope and to the sample via a 90/10 volume holographic beam splitter (Ondax). A 20x NIR objective (Zeiss) was used to focus the laser onto the sample. An excitation power of 5 mW was used, measured just before the objective. The backscattered Raman signal and Rayleigh reflection are collected by the objective and passed back to the beam splitter, where 90\% of the Rayleigh is eliminated and the Raman signal is transmitted. An achromatic half-wave plate is used to rotate the signal polarization back to the polarization of the laser before the monochromatic half-wave plate (parallel configuration) or perpendicular to it (perpendicular configuration). Another polarizer (called the analyzer) filters the polarization of the signal. The remaining Rayleigh component of the signal is then attenuated by two volume holographic notch filters (Ondax). The signal is then directed to a one meter spectrometer (Horiba) and detected by a 1D InGaAs detector array (Horiba). Due to fixed-pattern dark noise in the InGaAs detector, a dark spectrum is subtracted from each spectrum. We found the system could achieve an extinction ratio ($I_{\perp}/I_{\parallel}$) between 0.001 and 0.01 depending on the orientation of the 1st 1/2-$\lambda$ plate.

To accumulate a Raman crystallography dataset, we rotated the polarization orientation of the incident laser beam (of polarization $\vec{e}_{i}$, see Figure~\ref{ExpSetup}) stepwise from $\theta=0 - 360^{\circ}$ in a plane parallel to the $\alpha$-AgI (110) crystal face. The backscattered Raman signal is then filtered for light polarized parallel ($\vec{e}_{s,\parallel}$) and perpendicular ($\vec{e}_{s,\perp}$) to the incident light. At each $\theta$ step, a Raman spectrum is collected for both $\vec{e}_{s,\parallel}$ and $\vec{e}_{s,\perp}$ analyzer configurations. 
\medskip

\noindent{\textit{Fitting of PO Raman data}}

\begin{figure*}
	\centering
	\includegraphics[width=10 cm]{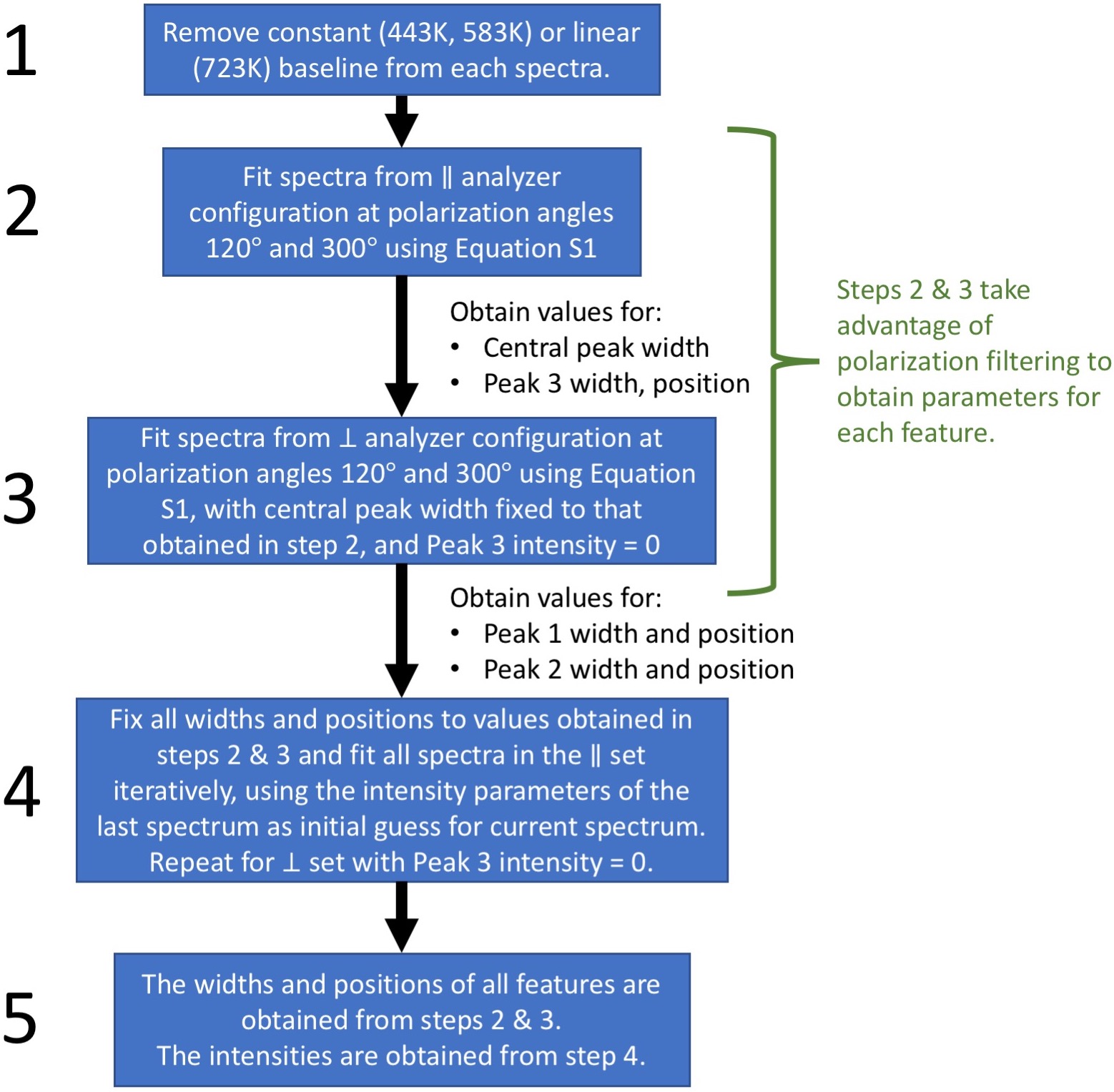}
	\caption{
		Flowchart explaining the fitting procedure of the PO Raman data employed to extract the positions, widths, and intensities of the features. 
	}
	\label{flowchart}
\end{figure*}

To aid the reader in following our PO fitting procedure, a flow chart is provided in Figure~\ref{flowchart}. 

To prepare the raw data for fitting, a small constant background was removed from each spectrum for measurements at 443 K and 583 K (step 1 Figure~\ref{flowchart}). This background results from fluctuations in the dark noise of the InGaAs detector. Measurements at 723 K showed a small but linear background with intensity increasing toward higher wavenumber. This background was removed with a linear fit to the starting and ending points of the data. This linear background may be due to blackbody radiation, which is more pronounced at 723 K, in addition to the dark noise fluctuations from the InGaAs detector.

We fit each spectrum in the PO Raman dataset with a sum of damped Lorentz oscillators, a single Debye relaxor, and a baseline correction ($<$2\% of total intensity). 
The Debye relaxor is used to fit the central component, according to previous treatments \cite{Yaffe:2017aa,Bouziane:2003aa,Fontana:1990aa}. 
A damped Lorentz oscillator was required to capture the extremely broad, low frequency features found in the spectra. 
While we believe that each peak is likely broadened due to a combination of lifetime and inhomogeneous broadening (arising from the many local environments of the `local tetrahedral oscillator'), we found that a Gaussian lineshape was un-physical in this low frequency range.
This is because the density of states of a Gaussian lineshape does not reach zero at zero frequency. Zero response at zero frequency is required for a purely oscillatory spectral element. 
The sum of oscillators was multiplied by the Bose-Einstein population factor ($ c_{BE}(\nu,T) $) to account for the temperature dependence of the vibrational state populations. 
The fitted equation is:
\begin{equation}
I(\nu,T)=b+c_{BE}(\nu,T) \left(\dfrac{c_{0} \gamma_{0} \nu}{\nu^{2} + \gamma_{0}^{2}} + \sum_{i=1}^{n} \dfrac{c_{i} \gamma_{i}^{3} \nu}{\nu^{2} \gamma_{i}^{2} + (\nu^{2}-\nu_{i}^{2})^{2}}\right)
\label{fiteqn}
\end{equation}
where $ I $ is the Raman intensity, $ b $ is a constant baseline correction, $ \nu $ is the Raman shift, the $ \nu_{i} $ are the resonance energies of the damped Lorentz oscillators, $\gamma_{0}$ and $\gamma_{i}$ are the damping coefficients of the Debye relaxor and Lorentz oscillators, $c_{0}$ and $c_{i}$ are the intensity coefficients of the Debye relaxor and Lorentz oscillators, and $ T $ is the temperature. 
Due to instrumental artifacts affecting the Stokes/Anti-Stokes ratio, we fit only the Stokes data with the temperature fixed to the sample temperature.

We fit the complete set of PO spectra using two guiding principles. 
	The first principle is that since all spectra in a Raman crystallography data set are taken under the same conditions of temperature, pressure, laser power, etc., all spectra reflect the same fundamental set of features whose widths and positions are fixed with respect to polarization orientation and analyzer configuration. 
	The second principle is that features with different symmetry respond differently to the orientation of the polarized light and the analyzer configuration, allowing isolation of particular features according to their symmetry.
 
\begin{figure*}
		\centering
		\includegraphics[width=7 cm]{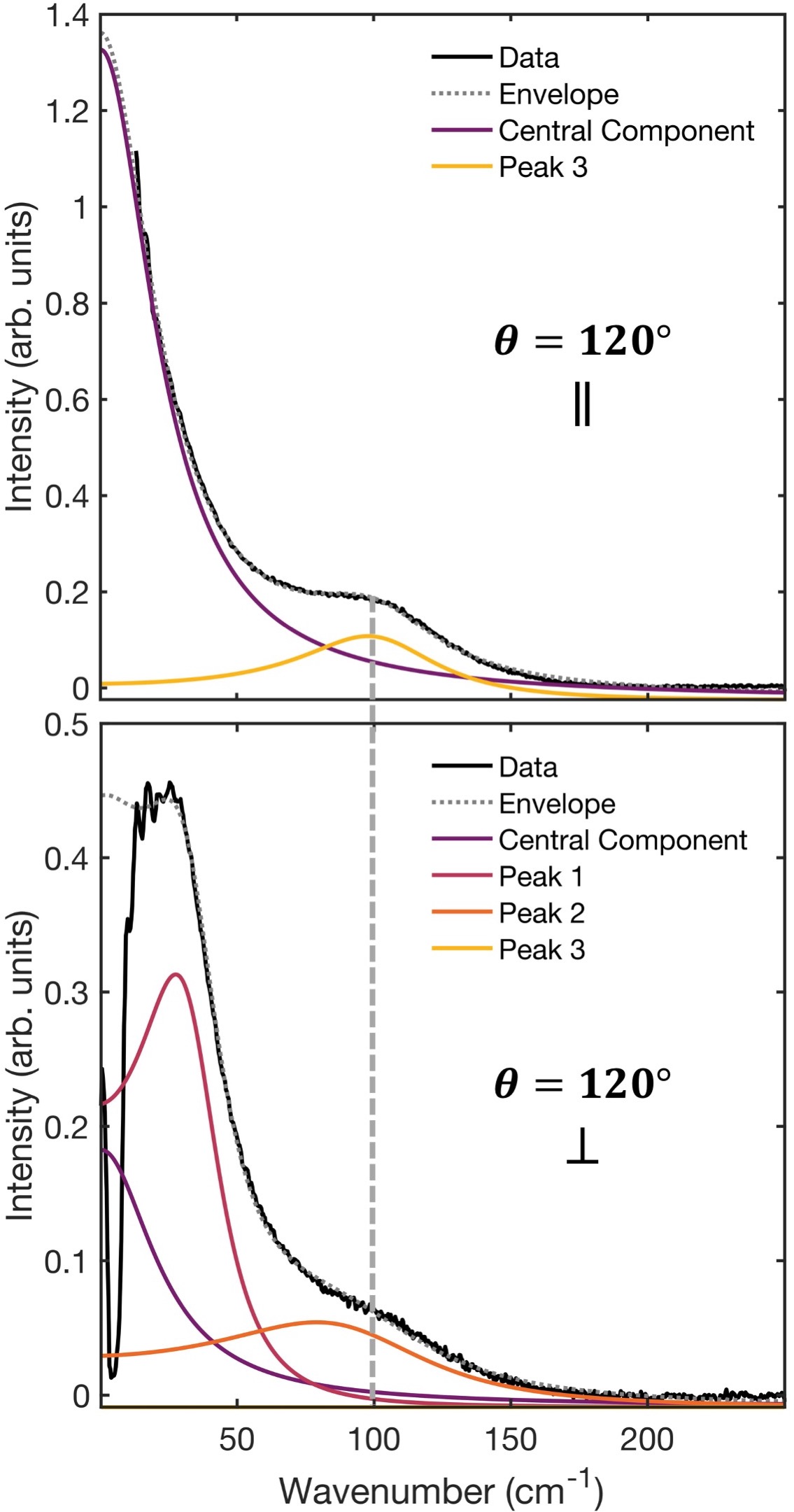}
		\caption{
			Fits to the $\parallel$ and $\perp$ Raman spectra at a polarization of $\theta=120^{\circ}$, used to determined the widths and positions of all four features. The gray dotted line demonstrates that Peaks 2 \& 3 have different widths and positions.
		}
		\label{prefit}
	\end{figure*}

\begin{figure*}
	\centering
	\includegraphics[width=18 cm]{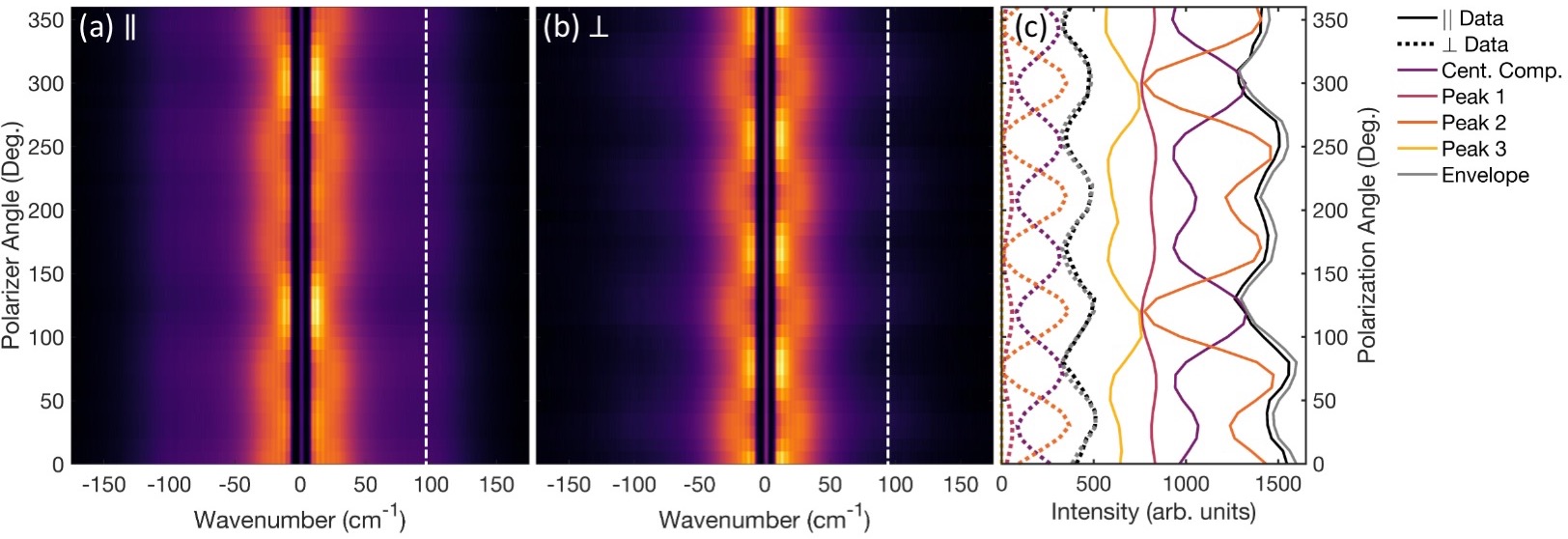}
	\caption{
		Deconvolution of the second shoulder of the Raman spectrum near 100 cm$^{-1}$ at 443K. (a) – (b) Intensity maps of the parallel ($ \parallel $) and perpendicular ($ \perp $) polarization configurations showing the location of a linecut at 90 cm$^{-1}$, centered on the second shoulder in the Raman spectrum. (c) Deconvolution of the linecuts of the $ \parallel $ and $ \perp $ PO Raman data at 90 cm$^{-1}$, as a function of polarizer angle. The deconvolution shows how the shoulder at 90 cm$^{-1}$ decomposes into the fundamental Raman features - the central component and Peaks 1-3. This plot shows that there is a nearly-constant component (Peak 3) in the $ \parallel $ configuration that is not present in the $ \perp $. The remaining intensity is captured by the central component and Peaks 1 \& 2. The parallel fit components have been shifted by the maximum value of Peak 3 for clarity.
	}
	\label{Linecut}
\end{figure*}

	We used the symmetry-filtering effects of the polarized light to obtain the widths and positions of each feature by performing pre-fits (steps 2 \& 3 in Figure~\ref{flowchart}) at the polarization angle of $\theta=120^{\circ}$ (Figure~\ref{prefit}). At this angle in the parallel configuration we found that only the central component and the shoulder at 100 cm$ ^{-1} $ are present. In perpendicular, the shoulders at 40 cm$ ^{-1} $ and 100 cm$ ^{-1} $ are maximized.
	These pre-fits revealed that the shoulder near 100 cm$ ^{-1} $ is actually composed of two peaks (Peaks 2 and 3, see Figure~\ref{prefit} and Figure~\ref{Linecut}). 
	This conclusion is based on the fact that the peak employed to fit this shoulder was found to have a different position and width in the parallel configuration than in perpendicular (Figure~\ref{prefit}). 
	This clearly indicates that this shoulder is composed of more than one feature, one that is expressed in parallel and another expressed in perpendicular at $\theta=120^{\circ}$.
	Therefore the fit to the perpendicular configuration yields the parameters for one feature (Peak 2) while the fit to parallel yields the parameters for the other feature (Peak 3). Fits to the complete PO dataset described below revealed that indeed Peak 2 \& 3 are distinct, having different symmetries (Figure 2c of main text).

The shoulder at 40 cm$ ^{-1} $ (Peak 1) proved difficult to fit uniquely. 
The width and position of this peak can be fit with a range of values that all yield equally good fits. We have chosen a value yielding good fits in both parallel and perpendicular, but we emphasize that there is significant uncertainty in this value. 
However, the polarization/orientation dependencies and symmetries reported in the main text are obtained regardless of the exact position and width of Peak 1, within the uncertainty window.

	Once width and position parameters for all four features were obtained from the prefits performed at 120$^{\circ}$ (steps 2 \& 3 in Figure~\ref{flowchart}), the full PO dataset could be fit to extract the dependence of intensity on polarization orientation for all features. 
	In this step (step 4 in Figure~\ref{flowchart}), we fixed all widths and positions to the values obtained in the pre-fits (Table~\ref{Fit_Params}). We then fit the spectra iteratively, beginning with 0$^{\circ}$, and using the intensity parameters from the fit of the preceding spectrum as an initial guess for the current spectrum. The parallel and perpendicular data sets were fit separately, but with the same set of widths and positions. We extracted from these fits the intensity coefficients of each component as a function of polarization angle as reported in Figure 2c of the main text (step 5 in Figure~\ref{flowchart}). We could not detect the presence of Peak 3 in the perpendicular dataset, and therefore its intensity was set to zero for all polarization angles.

\medskip

\noindent{\textit{Theory of PO Raman Symmetry Determination}}

In a single crystal, the intensity of a Raman mode $i$ ($S_{i}$) can vary with polarization orientation angle $\theta$ (see Figure~\ref{ExpSetup} above) as well as with the polarization configuration ($\parallel$ or $\perp$) according to the Raman polarizability tensor ($R_{i}^{j}$) of that mode, where $j$ accounts for the degeneracy of mode $i$. The Raman polarizability tensors for all point groups employed in our modeling of the Raman are given in Table~\ref{ramantensors} \cite{Hayes:2004aa}.
\begin{table}
	\caption{Raman polarizability tensors for each of the point groups used for the quantitative modeling of the Raman spectra presented in Figure~\ref{POModels} and Figure 2 of the main text \cite{Hayes:2004aa}. The $O_{h}$ point group applies to the space group model and the $D_{2d}$ point group applies to the `local tetrahedral oscillator' model, discussed below.}	
	\begin{ruledtabular}
\begin{tabular}{|c|c|c|c|c|c|c|} 
	\hline
	Point Group& \multicolumn{6}{c|}{Raman Tensors for Each Irreducible Representation ($R_{i,j}$)} \\ 
	\hline 
	Irreducible Representation&  $A_{1g}$&  \multicolumn{2}{c|}{$ E_{g} $}& \multicolumn{3}{c|}{$ T_{2g} $} \\ 
	\hline 
	$ O_{h} $&  
	$\left(\begin{array}{ccc}a& 0& 0\\ 0& a& 0\\ 0& 0& a\\\end{array}\right)$&  
	$\left(\begin{array}{ccc}b& 0& 0\\ 0& b& 0\\ 0& 0& -2b\\\end{array}\right)$& 
	$\left(\begin{array}{ccc}-\sqrt{3}b& 0& 0\\ 0& \sqrt{3}b& 0\\ 0& 0& 0\\\end{array}\right)$&  
	$\left(\begin{array}{ccc}0& d& 0\\ d& 0& 0\\ 0& 0& 0\\\end{array}\right)$&  
	$\left(\begin{array}{ccc}0& 0& d\\ 0& 0& 0\\ d& 0& 0\\\end{array}\right)$&  
	$\left(\begin{array}{ccc}0& 0& 0\\ 0& 0& d\\ 0& d& 0\\\end{array}\right)$\\ 
	\hline 
	Irreducible Representation&  \multicolumn{2}{c|}{$A_{1}$}&  $ B_{1} $&  $ B_{2}(z) $& \multicolumn{2}{c|}{$ E(x,y) $} \\ 
	\hline 
	$D_{2d}$& 
	\multicolumn{2}{c|}{$\left(\begin{array}{ccc}a& 0& 0\\ 0& a& 0\\ 0& 0& b\\\end{array}\right)$}&  
	$\left(\begin{array}{ccc}c& 0& 0\\ 0& -c& 0\\ 0& 0& 0\\\end{array}\right)$& 
	$\left(\begin{array}{ccc}0& d& 0\\ d& 0& 0\\ 0& 0& 0\\\end{array}\right)$& 
	$\left(\begin{array}{ccc}0& 0& e\\ 0& 0& 0\\ e& 0& 0\\\end{array}\right)$& 
	$\left(\begin{array}{ccc}0& 0& 0\\ 0& 0& e\\ 0& e& 0\\\end{array}\right)$\\
	\hline 
\end{tabular}
\end{ruledtabular}
\label{ramantensors}
\end{table} 
For the $\parallel$ and $\perp$ polarization configurations, respectively, the Raman intensity will be proportional to:
\begin{equation}
S_{i,\parallel}(\theta)=\sum_{j} \left(\vec{e}_{i}(\theta) \cdot R_{i}^{j} \cdot \vec{e}_{s,\parallel}(\theta)\right)^{2}
\label{intpar}
\end{equation}
\begin{equation}
S_{i,\perp}(\theta)=\sum_{j} \left(\vec{e}_{i}(\theta) \cdot R_{i}^{j} \cdot \vec{e}_{s,\perp}(\theta)\right)^{2}
\label{intperp}
\end{equation}
where the sum over $ j $ accounts for the degeneracy of the mode, and $ \vec{e}_{i} $, $\vec{e}_{s,\parallel}$, and $\vec{e}_{s,\perp}$ are the incident and scattered light (selected for parallel and perpendicular polarization configurations), respectively, as indicated in Figure~\ref{ExpSetup} \cite{Mizoguchi:1989aa}. 

The symmetry of a particular mode can be identified from the periodicities in its intensity response ($S_{i}(\theta)$), making Raman crystallography a powerful tool in the identification of modes. Additionally, the crystal orientation can be determined from PO Raman (see below).
Two other features of $S_{i,\parallel}(\theta)$ and $S_{i,\perp}(\theta)$ can serve to confirm mode assignment. First, all modes share the same phase in $\theta$, meaning if a particular mode requires a phase shift $\Delta$ (i.e. $S_{i,\parallel}(\theta + \Delta)$, $S_{i,\perp}(\theta + \Delta)$) compared to the other modes to fit the data, then that mode has been incorrectly identified. Second, the ratio $S_{i,\perp}(\theta)/S_{i,\parallel}(\theta)$ must also agree with the theoretical calculation, meaning the intensity in the perpendicular must be at the proper ratio compared to the parallel. This serves as a check on both mode identification and the quality of the optical alignment. Both the phase and intensity checks have been confirmed for the mode assignment presented in this manuscript.

Equations~\ref{intpar} \& \ref{intperp} have been employed to calculate the PO response of the \hmn{Im-3m} ($ O_{h}^{9} $) space group model of $\alpha$-AgI as shown in Figure~\ref{POModels}. A modification of Equations~\ref{intpar} \& \ref{intperp} (Equations~\ref{intpard2d} \& \ref{intperpd2d}) is employed below in the `local tetrahedral oscillator' model of the PO response presented in the main text (Figure 2).
\medskip

\noindent{\textit{Self-Consistent Determination of $\alpha$-AgI (110) Orientation}}

\begin{figure*}
	\centering
	\includegraphics[width=18 cm]{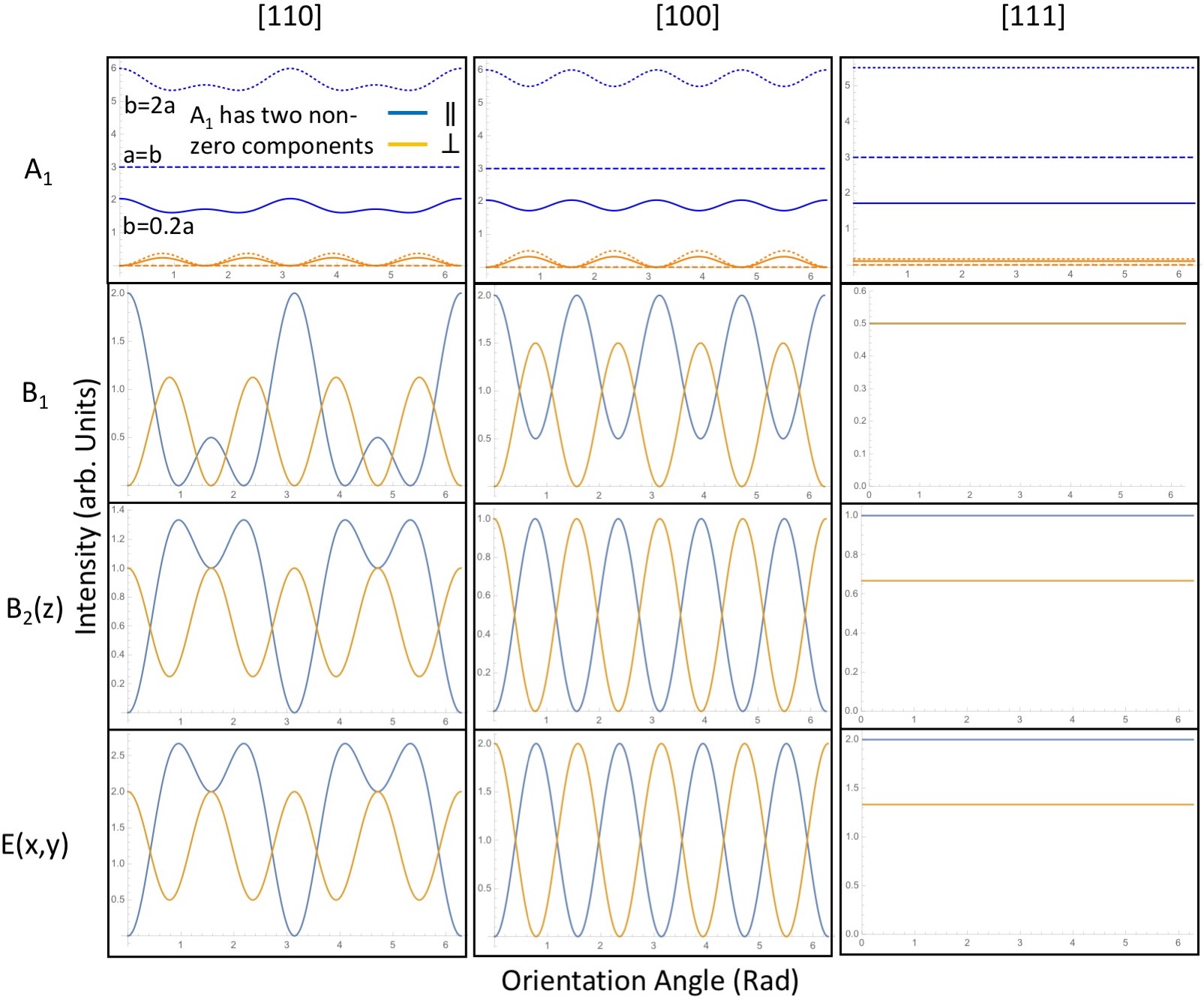}
	\caption{
		Intensity response as a function of polarization orientation angle for the four possible Raman mode symmetries of the $D_{2d}$ `local tetrahedral oscillator', for the three highest symmetry directions of the BCC cubic unit cell. The parallel ($\parallel$, blue) and perpendicular ($\perp$, yellow) responses are plotted together. 
	}
	\label{orient}
\end{figure*}

Using Equation~\ref{intpard2d} and Equation~\ref{intperpd2d} discussed in the `local tetrahedral oscillator' model below and the experimental configuration of Figure~\ref{ExpSetup}, we have calculated the expected intensity response of each possible Raman mode symmetry as a function of polarization orientation angle for the three possible high symmetry faces of the crystal, namely the $ \{100\} $, $\{110\}$, and $\{111\}$ classes of planes. The results are shown in Figure~\ref{orient}.
 The $\{111\}$ planes show no orientation dependence for all modes. The $\{100\}$ planes show only 90$^{\circ}$ periodicity. Only the $\{110\}$ planes show the appropriate combination of 180$^{\circ}$ and 90$^{\circ}$ periodicity in the parallel configuration and 90$^{\circ}$ periodicity in the perpendicular, matching with the experimental findings in the main text. Thus, the $\{110\}$ planes provide the best choice of facet to explain the experimental results. This finding is consistent with the expectations of literature for preparing an \AgI\ crystal from a $\beta$-AgI single crystal \cite{Mills:1970aa}.
\medskip

\noindent{\textit{Molecular Dynamics Simulations}}

\begin{figure*}[]
	\centering
	\includegraphics[width=8.8 cm]{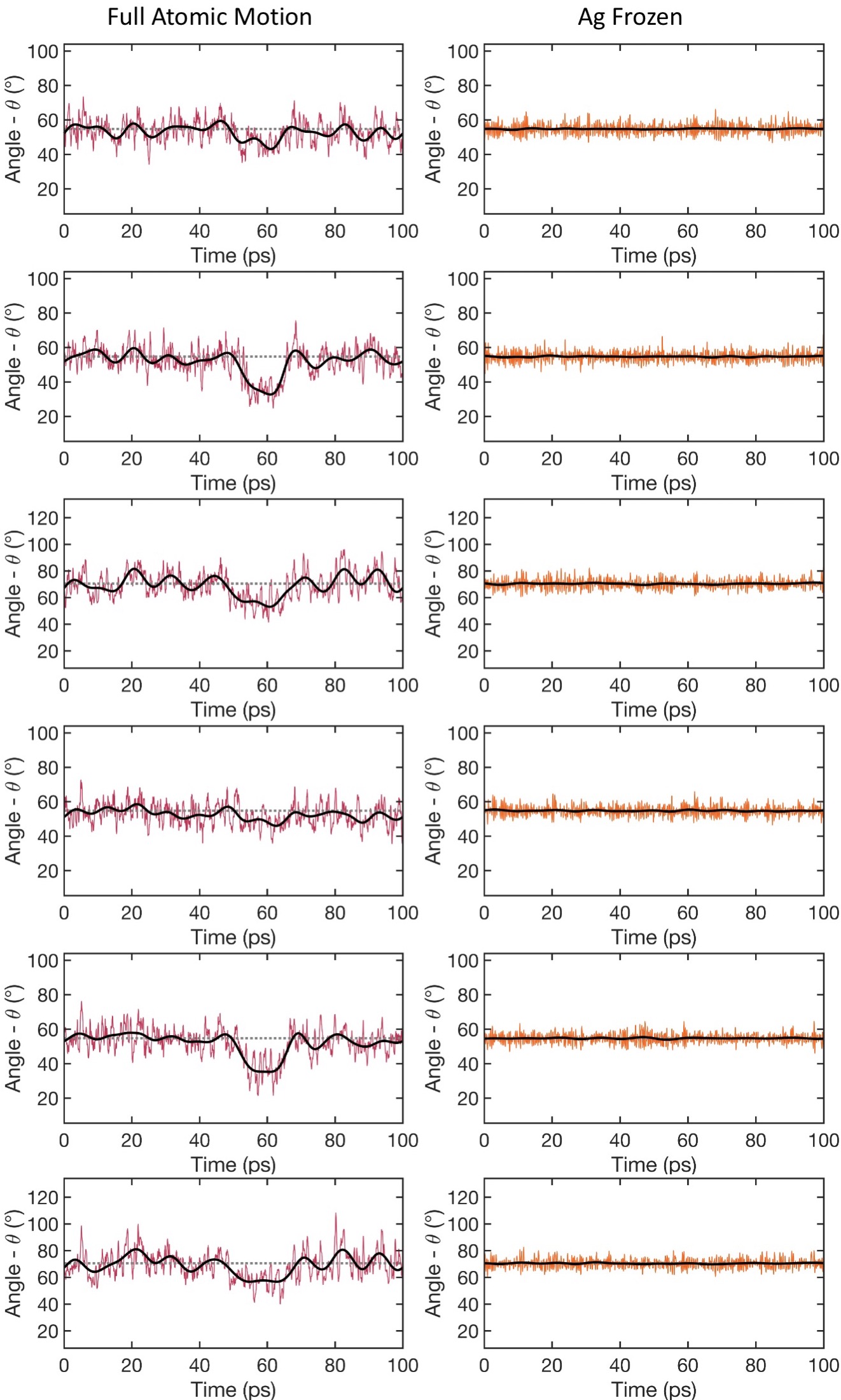}
%
	\includegraphics[width=8.8 cm]{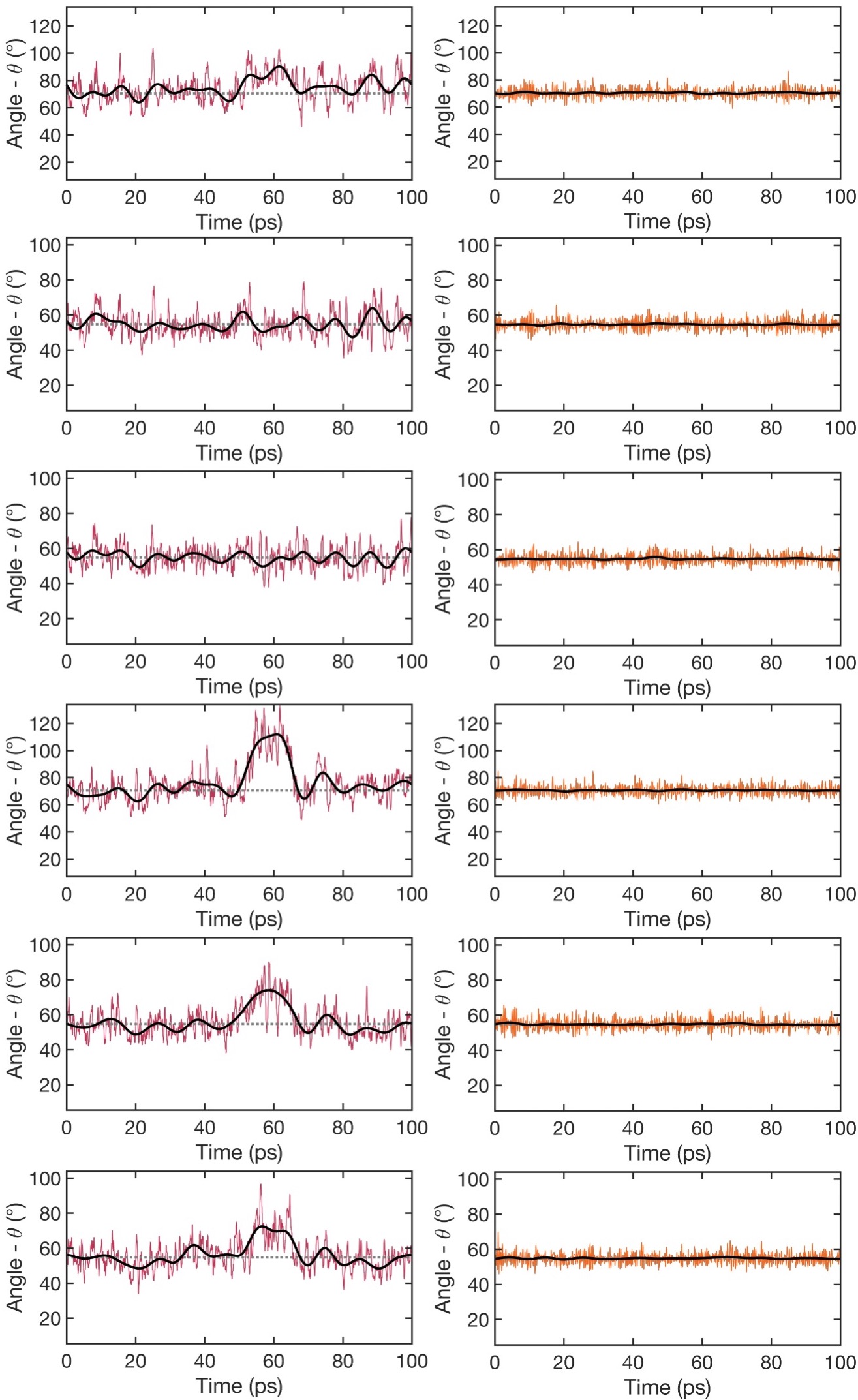}
	\caption{
		All 12 I-I-I tetrahedral angles for the selected tetrahedron analyzed in the \textit{ab-initio} molecular dynamics simulation. Full Ag$^{+}$ motion shown in 1\textsuperscript{st} \& 3\textsuperscript{rd} columns. Frozen Ag$^{+}$ simulation shown in 2\textsuperscript{nd} \& 4\textsuperscript{th} columns.  
	}
	\label{anglesfig}
\end{figure*}

First-principles (\textit{ab-initio}) molecular dynamics (AIMD) simulations were performed using the VASP code \cite{Kresse:1996aa}. The projector augmented wave (PAW) method was used to treat core-valence interactions \cite{Kresse:1999} .  Exchange-correlation was described with the PBE functional \cite{Perdew:1996}. An energy threshold of 10$^{-6}$~eV, a single k-point, and a plane-wave cutoff energy set to 280~eV were used.
As a timestep in solving the classical equations of motions we applied 10~fs. A canonical (NVT) ensemble with 128 atoms in a unit cell of volume of {\raise.17ex\hbox{$\scriptstyle\mathtt{\sim}$}}4736 \AA$^{3}$ was simulated at 500K: the system was first equilibrated for 10~ps, followed by a 100~ps production run.  
The mean squared displacement, $ MSD $, was calculated as 
\begin{equation}
 MSD(t) = \langle \left( \vec{R}(t) - \vec{R}(0) \right)^2 \rangle ,
\end{equation}
where $\langle ... \rangle$ denotes the average over atoms and  $\vec{R}(t)$ is the position of an atom at time $t$.
From it the diffusion constant can be obtained via: $ MSD(t)=6 D t $. To calculate the tetrahedral angles ($\Theta_{i}$), first the I atoms forming one AgI$_{4}$ tetrahedron were identified. Then, we calculated atomic distances $ \vec{d}_{ij} (t)=\vec{R}_{i} (t) - \vec{R}_{j} (t) $ from the trajectories $ \vec{R}_{i} (t) $, where the indices $i$ and $j$ address the I atoms of the tetrahedron. Finally, from the atomic distances, the angles have been calculated as:
\begin{equation}
\Theta_{i}=\arccos \left(\dfrac{\vec{d}_{ij} \cdot \vec{d}_{ik}}{\lvert \vec{d}_{ij} \rvert \lvert \vec{d}_{ik} \rvert}\right)
\label{angles}
\end{equation}
where  $i$, $j$ and $k$ address I atoms of the tetrahedron. From the 12 possible angles, we selected one representative case and showed it in the main text, since the qualitative behavior of all angles was similar (see Figure~\ref{anglesfig}).

For the \textit{gedanken-experiments} discussed in the main text, either the I or Ag atoms were frozen by using selective dynamics as implemented in VASP. 
After this freezing, the system was equilibrated again for 10 ps. The thermostat for the frozen systems was set such that the temperature associated with the available degrees of freedom was set to 500 K. For the cases in which Ag atoms were selected from MD snapshots, 10 instantaneous Ag configurations, separated by 1 ps along the trajectory, were chosen from the MD run.

In our analysis, we have Fourier-filtered the nuclear trajectories according to
\begin{equation}
R_{filtered}(t) = F^{-1} \left[ \Theta(\omega) \Theta(\omega_{max} - \omega) F[R(t)] \right]
\label{fourier}
\end{equation}
where $F$ denotes the Fourier transform and $\Theta$ the Heaviside step function. The upper limit in the frequency, $\omega_{max}$, was set to 4 cm$^{-1}$. 

The frequency dependent vibrational density of states, $ VDOS(\omega) $, was calculated as the Fourier transform of the normalized velocity autocorrelation function $VACF (t)$: 
\begin{equation}
VDOS(\omega) = \frac{N}{V} \  F \left[ \frac{VACF(t)}{VACF(0)} \right] 
\label{vacfeqn}
\end{equation}
where $N$ is the number of contributing atoms and $V$ the simulation volume.
The velocity autocorrelation function, $ VACF(t)$, was calculated as:
\begin{equation}
VACF(t) = \langle \nu(\tau + t) \cdot \nu(\tau) \rangle
\label{vacfeqn}
\end{equation}
where $\langle ... \rangle$ denotes the average over atoms and initial time, $\tau$, and $\nu(t)$ denotes the velocity of an atom at time $t$. From this, the diffusion constant, $D$, can be determined as:
\begin{equation}
D= F \left[VACF(t) \right] (\omega=0)/6	
\end{equation}


\newpage
\section{The `Local Tetrahedral Oscillator' Model}

Our model is motivated by an analysis of whether a phonon picture of vibrations can extend to the vibrations of mobile Ag$^{+}$ in $\alpha$-AgI. To propagate a phonon in this system, a stable ordered state of atoms is required, and a correlation length for the phonon is established by the residence time of Ag$^{+}$ ($ \tau \approx 2-3 $ ps) \cite{Funke:1976aa,Nemanich:1980aa,Boyce:1977aa,Salamon:1979aa}. The decay time for Ag$^{+}$ ordering at the unit cell level, assuming four Ag atoms in the unit cell that hop independently, is $ \tau \approx 500 $ fs. Ordering beyond the unit cell has an even shorter decay time. While correlated Ag$^{+}$ motion may increase this value somewhat, it is clear from this estimate that crystallographic order does not exist in this material on the time scales required for substantial phonon propagation. A phonon-like picture of vibrations is therefore unlikely to be useful, and a local view is needed.

Considering the above, we propose the vibrational dynamics to be localized to a sub-unit cell structure. As the Ag$^{+}$ are tetrahedrally coordinated by I$^{-}$, it is natural to consider a tetrahedral model for the vibrational dynamics. As stated in the main text, we propose the fundamental unit of vibration is the orientationally ordered AgI$_{4}$ tetrahedron. We note that other tetrahedral structures, like IAg$_{4}$, could also serve as the fundamental unit of vibration, and more work is needed to discern exactly what structure is observed. When Ag$^{+}$ hops, one tetrahedral unit is lost followed by the formation of a new unit at the new site. We assume these tetrahedra vibrate independently, being localized by the strong dynamic structural disorder. A similar tetrahedral model was proposed to explain the Raman spectrum of RbAg$_{4}$I$_{5}$, but the large number of tetrahedral orientations in this structure prevented a full assignment of mode symmetries \cite{Gallagher:1979aa}. We now derive the Raman response presented in Figure 2c of the main text, Table~\ref{Irreps}, Table~\ref{Assignments}, and Figure~\ref{POModels} from this conceptual model.

In our `local tetrahedral oscillator' model, $\alpha$-AgI does not possess a space group because of the lack of translational symmetry of Ag$^{+}$ discussed above. However, the symmetries observed in optical vibrations depend only on the point group, which does not include translational symmetries. Only a preservation of orientational symmetry is required to detect point group symmetries in our PO measurement, and this is preserved by the iodine sublattice. Therefore, we employ a point group model for the Raman mode symmetries. Tetrahedral configurations in a BCC lattice have $D_{2d}$ point group symmetry. This symmetry is not cubic and is the point group for certain tetragonal space groups. The tetrahedron is distorted from the ideal, isotropic $T_{d}$ tetrahedral point group. We have confirmed from our MD simulations (Fig. 3b in the main text) that the time-average structure retains the $D_{2d}$ symmetry and we therefore employ this point group in our model. 

\begin{figure*}
	\centering
	\includegraphics[width=8 cm]{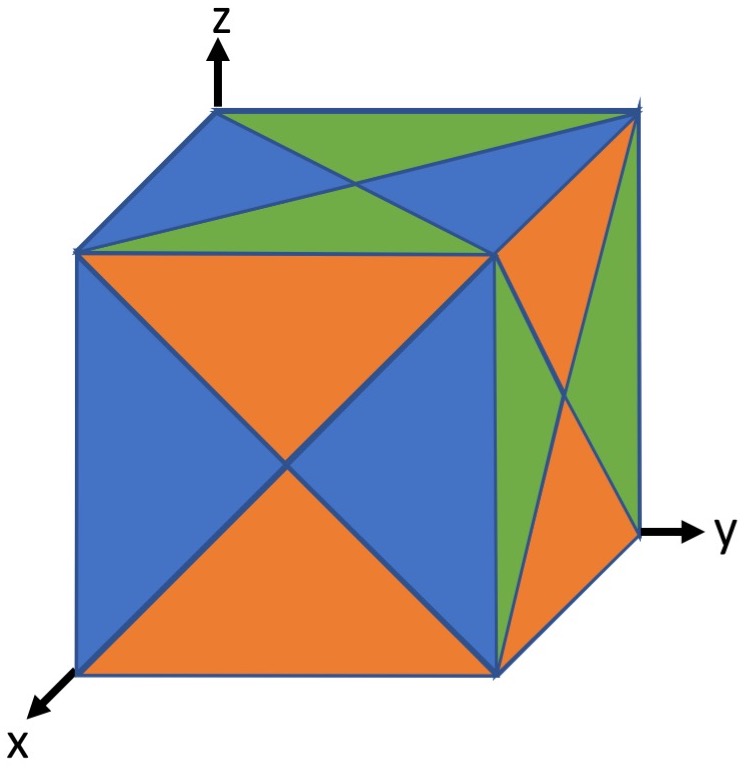}
	\caption{
		Inequivalent tetrahedral orientations in the BCC unit cell. Each triangle on the face of the cube represents a $D_{2d}$ tetrahedral site. There are 24 in total within the BCC unit cell. There are only 3 inequivalent sets when considering the Raman response. These three sets are colored with different colors (orange, green, and blue) on the surface of the cube. Each type appears with equal frequency. The opposite (non-visible) sides of the cube have the same color pattern as their corresponding visible side.   
	}
	\label{bcctetrahedra}
\end{figure*}

We assume that at any given time, the crystal is inhabited by many $D_{2d}$ `local tetrahedral oscillators' occupying the tetrahedral sites, of which there are several different orientations within the crystal. We expect the overall PO Raman response to be an average over all those orientations. To calculate the PO Raman response, we first calculate the PO response for each of the irreducible representations of the $D_{2d}$ point group from their Raman polarizability tensors (Table~\ref{ramantensors}) \cite{Hayes:2004aa} and Equations~\ref{intpar} and \ref{intperp}, for each of the possible orientations of the tetrahedron in the BCC structure. Symmetry considerations reduce the number of required calculations: there are only 3 inequivalent orientations within the BCC structure, as shown in Figure~\ref{bcctetrahedra}. 

We then sum the response over all orientations in the BCC unit cell, with equal weight on each orientation. The resulting response is:
\begin{equation}
S_{i,\parallel}^{D_{2d}}(\theta)=\sum_{k} \sum_{j} \left(\vec{e}_{i}(\theta) \cdot \Phi_{k}^{T} \cdot R_{i}^{j} \cdot \Phi_{k} \cdot \vec{e}_{s,\parallel}(\theta)\right)^{2}
\label{intpard2d}
\end{equation}
\begin{equation}
S_{i,\perp}^{D_{2d}}(\theta)=\sum_{k} \sum_{j} \left(\vec{e}_{i}(\theta) \cdot \Phi_{k}^{T} \cdot R_{i}^{j} \cdot \Phi_{k} \cdot \vec{e}_{s,\perp}(\theta)\right)^{2}
\label{intperpd2d}
\end{equation}
where $ R_{i}^{j} $ is the Raman polarizability tensor (given in Table~\ref{ramantensors}) of the $j$th degenerate component of the mode $i$, $ k = 1-3 $ is an index representing the three inequivalent orientations, and $\Phi_{k}$ is a rotation matrix that rotates $ R_{i}^{j} $ to the $k$th inequivalent orientation. The $ S_{i}^{D_{2d}} $ were calculated according to the experimental geometry shown in Figure~\ref{ExpSetup}, with $\vec{e}_{i}$, $\vec{e}_{s,\parallel}$, and $\vec{e}_{s,\perp}$ all rotating in the (110) plane of the BCC unit cell. The results of this calculation are presented in Figure~\ref{POModels} and are employed in Figure~2c of the main text to identify the symmetry of each Raman feature. The principal axis of the $D_{2d}$ tetrahedron aligns with the BCC c-axis in the unrotated configuration, while the remaining axes align with the a- and b- axes in the conventional Raman polarizability tensor.

The average over orientations generates a polarization dependence such that modes that had distinguishable symmetries no longer do (such as $B_{2}$ and $E$, see Figure~\ref{POModels}). These modes, while having the same effective symmetry, are still expected to be non-degenerate and should have different energies. However, we did not detect this in our measurements, and this lead us to the assignment in Table~\ref{Assignments} of two irreducible representations to one feature. One reason that these modes cannot be distinguished by energy may be inhomogeneous broadening due to the many possible local environments of the `local tetrahedral oscillators'. The tetrahedral relaxational motions observed in Figure 3 in the main text, which generate long-lived distortions from the perfect $D_{2d}$ structure, as well as changing configurations of nearby Ag$^{+}$ ions may smear out the oscillation frequencies of each mode, leading to overlap of energies and the overall broad features we see. 

The average over orientations also generates a polarization dependence for the $B_{1}$, $B_{2}(z)$, and $E(x,y)$ modes that resembles a higher tetrahedral symmetry, namely $T_{d}$. This, combined with the fact that we could not separate some pairs of modes by energy, gave the impression that the tetrahedral structure may be of higher symmetry than expected. 
However, we did detect one feature distinguishing the $D_{2d}$ model from the $T_{d}$: the polarization dependence of the $A_{1}$ mode. In the $T_{d}$ model, no polarization dependence is expected for the $A_{1}$. 
However, in the $D_{2d}$ model a polarization dependence is expected (Figure~\ref{POModels}) even after the averaging over orientations. 
The PO dependence is determined by the values of the tensor components $a$ and $b$ in Table~\ref{ramantensors}. 
Peak 3 has a polarization dependence matching $A_{1}$ symmetry under the $D_{2d}$ model. 
We made this identification by fitting the intensity response of Peak 3 in the parallel configuration  with Equation~\ref{intpard2d} (Figure 2c of the main text) to obtain the ratio $a/b$. The values are 0.38 at 443K, 0.52 at 583K, and 0.6 at 723K. Thus, the presence of polarization dependence in Peak 3 confirms the $D_{2d}$ model over the $T_{d}$.


\bibliography{20191128_AgI_Paper_References}